\documentclass[aps,a4,twocolumn]{revtex4-1}
\usepackage{amsmath}
\usepackage{latexsym}
\usepackage{float}
\usepackage{amssymb}
\usepackage{graphicx}
\usepackage{epsfig}

\newcommand{\be}{\begin{equation}}
\newcommand{\ee}{\end{equation}}
\newcommand{\ba}{\begin{array}}
\newcommand{\ea}{\end{array}}

\begin{document}
\title{A review of Monte Carlo simulations of polymers with PERM}



\author{Hsiao-Ping Hsu$^1$
and Peter Grassberger$^{2,3}$
}


\affiliation{$^1$Institut f\"ur Physik, Johannes 
Gutenberg-Universit\"at, Mainz, D-55099, Germany\\
$^2$Forschungszentrum J\"ulich, J\"ulich, D-52425 Germany \\
$^3$Complexity Science Group, University of Calgary, Calgary T2N 1N4, Canada}


\begin{abstract}

In this review, we describe applications of the pruned-enriched Rosenbluth method
(PERM), a sequential Monte Carlo algorithm with resampling, to various problems
in polymer physics. PERM produces samples according to any given prescribed 
weight distribution, by growing configurations step by step with controlled bias, and correcting
``bad" configurations by ``population control". The latter is implemented, in 
contrast to other population based algorithms like e.g. genetic algorithms, by 
depth-first recursion which avoids storing all members of the population at the 
same time in computer memory. The problems we discuss all concern single polymers
(with one exception),
but under various conditions: Homopolymers in good solvents and at the $\Theta$
point, semi-stiff polymers, polymers in confining geometries, stretched polymers 
undergoing a forced globule-linear transition, star polymers, bottle brushes,
lattice animals as a model for randomly branched polymers, DNA melting, and finally 
-- as the only system at low temperatures, lattice heteropolymers as simple models for 
protein folding. PERM is for some of these problems the method of choice, but it 
can also fail. We discuss how to recognize when a result is reliable, and we 
discuss also some types of bias that can be crucial in guiding the growth into 
the right directions. 

\end{abstract}

\maketitle
\section{Introduction}
\label{Intro}
  Research in the field of polymer physics has grown vigorously since the 
1950s~\cite{Flory,Lifshitz,deGennes,Grosberg}. Recent developments in 
the techniques for the tools of atomic force microscopy (AFM)~\cite{Roiter}, 
in fabrication of nanoscale devices and in single-chain manipulation 
techniques~\cite{Salman,Meller,Kasianowicz} open possibilities for a broad 
range of applications in physical chemistry, biotechnology and material science.
During this time, much effort has also been put into studying the 
statistical properties of polymers by computer simulations~\cite{Binder,Binder08}. 
Indeed, due to the richness of the observed phenomena and the non-triviality
of the problems involved, polymer physics has from the very beginning served
as a playground for developing novel Monte Carlo 
strategies~\cite{RR,Wall-Erpenbeck,Madras-Sokal}. These strategies depend 
strongly on the problems one is interested in: Linear {\it versus} branched 
polymers, dilute {\it versus} dense systems, scaling laws {\it versus} detailed
material properties, classical {\it versus} quantum mechanical problems, 
implicit {\it versus} explicit treatment of solvent, etc.

In this review we shall only deal with one class of algorithms, the 
pruned-enriched Rosenbluth method (PERM)~\cite{Grassberger97}.  So far it has 
been used for classical physics only, although closely related methods have also 
been used since long ago for quantum mechanical simulations~\cite{Anderson}. 
Although it is not a panacea and fails miserably in many problems, it still found 
applications to several of the above dichotoma, and in some cases it beats
the (presently known) competitors by huge margins.

In the following we shall mostly be concerned with single unbranched
molecules moving freely in a dilute solvent. Later we will also consider
branched polymers and polymers attached to surfaces.
  The basic characteristics of linear polymer chains depend on the
solvent conditions. At high temperatures or in good solvents repulsive 
interactions (the excluded volume effect) and entropic effects dominate the 
conformation, and the polymer chain tends to swell to a random coil.  
At low temperatures or in poor solvents, however, attractive interactions between 
monomers dominate the conformation and the polymer chain tends to collapse 
and form a compact dense globule. The coil-globule transition
point is called the $\Theta$-point. Based on field theory~\cite{deGennes}, 
the behavior of polymer chains in good solvents is well understood. 
In the thermodynamic limit (as the chain length $N \rightarrow \infty$), 
the partition function scales as 
\be
      Z \sim \mu_\infty^{-N} N^{\gamma-1} \enspace {\rm at}
\enspace T>T_\Theta
\label{eq-ZN-good}
\ee
where $\mu_\infty$ is the critical fugacity and $\gamma$ is 
the entropic exponent related to the topology. Below the $\Theta$-point,
a collapsed polymer can essentially be viewed as a liquid droplet. 
According to the Lifshitz mean field theory~\cite{Lifshitz,Grosberg},
a surface term should be included in the partition sum as
\be
     Z \sim a^N b^{N^s} N^{\gamma-1} \enspace {\rm at}
\enspace T<T_\Theta                     \label{Z-}
\ee
with $s=(d-1)/d$ and $b>1$.

Generally speaking, the thermodynamic limit of a polymer system coincides with
the limit when 
the chain length $N$ tends to infinity. For conventional Monte Carlo (MC) 
methods such as the Metropolis algorithm, one can only simulate moderately large 
systems, the maximal feasible values of $N$ depending on the temperature and on 
the degree of reality of the model. Going from simple lattice-based models at 
high temperatures to models with realistic interactions and further to folded 
proteins with explicitly included solvent, $N_{\rm max}$ might decrease from 
$10^4$ to $\ll 100$. If one is interested mostly in scaling laws (as we shall be),
one simulates at several values of $N$ and uses finite-size scaling (FSS) 
to extrapolate the behavior of the considered thermodynamic quantities
to $N \rightarrow \infty$. Rather often, either very large finite-size
effects have to be considered or it is too difficult to reach equilibrium states
or to produce sufficiently many independent configurations.  
For some problems (not for all!), it was a big breakthrough when 
(PERM)~\cite{Grassberger97,Grassberger98,GN00,Grassberger02} was proposed 
in 1997. It is particularly efficient for temperatures near the $\Theta$ collapse,
where chains of length up to $N=1,000,000$ could be sampled with high statistics, 
and it was confirmed unambiguously that the $\Theta$ collapse is a tricritical
phenomenon with upper critical dimension $d_c=3$~\cite{deGennes}.
Since then, many other applications have also been made. 
Many other applications have
also been made successfully by PERM~\cite{Grassberger02}, which provide in some cases 
a deep understanding on the scaling behavior of polymer chains
under different solvent conditions, geometrical confinements, on the phase
transition behavior of polymer chains adsorbed onto a wall, on polymers
stretched by a force, etc.

In the next section we give a detailed description of the basic algorithm. This 
algorithm can be made substantially more efficient by a suitable bias in the growth 
direction, and two biases (including `Markovian anticipation') are discussed in
Sect.~3. Applications are treated in Sects. 4 ($\Theta$-polymers), 5 (stretched polymers 
in poor solvents), 6 (semiflexible polymers), 7 (polymers in confining geometries),
8 (branched polymers with fixed tree topologies), 9 (lattice animals), 10 (protein 
folding), and 11 (DNA melting). Finally the paper concludes with a summary in Sect.~12.


\begin{figure*}[htb]
\begin{center}
\includegraphics[angle=0,width=0.90\textwidth]{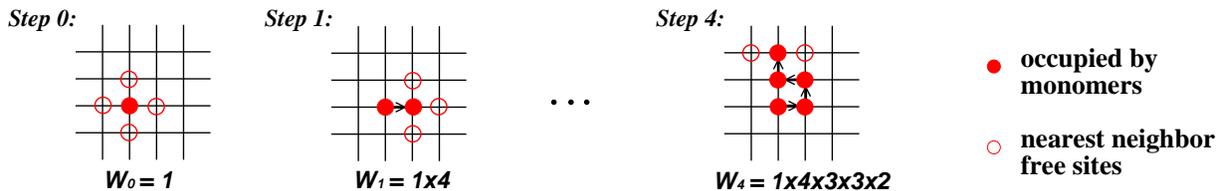}
\caption{Schematic drawings of building a SAW from the
$0^{\rm th}$ step to the $4^{\rm th}$ step and the associated
weight at each step. At the $0^{\rm th}$ step, we set the weight
$W_0=w_0=1$, and the probability $p_0=1$. The
Rosenbluth bias is used here such that $p_n=1/n_{\rm free}$
at each step, so the total weight $W_n=\Pi_{n=0}^{n=4} n_{\rm free}$.}
\label{fig-weight}
\end{center}
\end{figure*}

\section{Algorithm: Pruned-enriched Rosenbluth Method (PERM)}
\label{sec-PERM}
In statistical thermodynamics, the partition function for a canonical 
ensemble in thermal equilibrium is defined
by
\be
    Z(\beta)=\sum_{\alpha} Q(\alpha)=\sum_{\alpha}\exp(-\beta E(\alpha))
\ee
here $\beta=1/k_BT$, $E(\alpha)$ is the corresponding
energy for the $\alpha^{\bf th}$ configuration, $Q(\alpha)/Z$ is the 
Gibbs-Boltzmann distribution, and $Q(\alpha)=\exp(-\beta E(\alpha))$ is normally called the 
Boltzmann weight.  If each configuration is repeatedly and independently 
chosen according to a randomly chosen probability $p(\alpha)$ (a bias), 
the partition sum is rewritten as
\be
    Z = \lim_{M\to\infty} \hat{Z}     \label{eq-hatZ}
\ee
where $M$ is the number of trials and 
\be
    \hat{Z}=\frac{1}{M}\sum_{\alpha=1}^M Q(\alpha)/p(\alpha)
  =\frac{1}{M} \sum_{\alpha=1}^M W(\alpha) \;, 
\label{ZM}
\ee  
with modified weights 
\be
    W(\alpha)=Q(\alpha)/p(\alpha).
\ee 
If we use $p(\alpha) \propto \exp(-\beta E(\alpha))$ \{Gibbs sampling\}, each 
contribution to $\hat{Z}_M$ has the same weight, which is an example
of the so called 
`importance sampling'. The estimate for any observable $A$ is given by 
\be 
    \langle A \rangle  = \lim_{M \to \infty} \langle A \rangle_M = \lim_{M \to \infty} 
\frac{\sum_{\alpha=1}^M A(\alpha) W(\alpha)}
{\sum_{\alpha=1}^M W(\alpha)} \;. 
\ee
In general, we expect that statistical fluctuations of $\langle A \rangle_M$
are minimal, at given $M$, if we use importance sampling and if all
trials are independent. In general this is infeasible. The Metropolis
method achieves perfect importance sampling at the cost of highly 
correlated trials. PERM tries, with a completely different strategy, at a 
compromise between importance sampling and independence.

Things are best illustrated by a linear chain of $N+1$ monomers in an implicit 
solvent, modeled by an interacting self-avoiding walk (ISAW) of $N$ steps
on a simple (hyper-)cubic lattice of dimension $d$. The interactions in
this model are (i) the chain connectivity which enforces that adjacent 
monomers sit on adjacent lattice sites; (ii) self-avoidance that 
excludes configurations in which the same lattice site is occupied by 
two or more monomers; and attractive interactions (energies $\epsilon<0$) 
between non-bonded monomers occupying neighboring lattice sites. 
Writing $q=\exp(-\beta \epsilon)$ for the Boltzmann factor, the 
partition sum is 
\be 
     Z_N(q)=\sum_{\rm walks}q^m 
\label{eq-ZNSAW}
\ee
where $m$ denotes the total number of non-bonded nearest neighbor pairs.
The solvent quality is 
varied by changing the temperature $T$. 

In the original Rosenbluth-Rosenbluth (RR) method~\cite{RR}, polymer chains
are built like random walks by adding one monomer at each step.
At the $0^{\rm th}$ step, the first monomer is placed at
an arbitrary lattice site. For this ``chain" of length $N=0$, the weight 
is trivially $W_0=1$. For the first step one has $2d$ possibilities and 
no interactions yet, giving $W_1 = 2d$. For subsequent steps one has to 
take self-avoidance into account. When a monomer is added to a chain of 
length $N-1$, one scans the neighborhood of the chain end to identify the
free sites on which a monomer could be added. If there are $n_{\rm free}\geq 1$
free neighbors, the next step is chosen uniformly among them, while the 
walk is killed if $n_{\rm free}=0$ (``attrition"). After this step the 
weight $W_N$ is updated by multiplying $W_{N-1}$ by 
\be
   w_n=q^{m_n}/p_n,      \label{wn}
\ee
where $p_n= 1/n_{\rm free}$ and $m_n$ is the number of neighbors of the new 
site already occupied by non-bonded monomers (notice that $m=\sum_{n=0}^{N}m_n$).
Therefore, after $N$ steps the total weight is 
\be
     W_N = w_NW_{N-1}= \ldots = w_Nw_{N-1} \dots w_0=\prod_{n=0}^N w_n  \;.
\label{Wn}
\ee
When the chain length $N$ becomes very large, the method fails for two reasons:
First of all, attrition implies that only an exponentially small fraction of 
trials survive and give any contribution at all. Secondly, as the weight
factors $w_n$ are weakly correlated random variables, the full weight $W_N$
will show huge fluctuations. Thus the surviving configurations will finally
be dominated by a single configuration, demonstrating a dramatic lack of 
importance sampling.

PERM~\cite{Grassberger97} was invented to overcome this shortcoming of the
RR method. The main spirit of PERM is as follows,
\begin{itemize}
\item Polymer chains are built like random walks by adding one
monomer at each step.
\item A Rosenbluth-like bias is used for choosing one of the 
nearest neighbor free sites for the next step of the walk, but
a wide range of probability distributions
can be used depending on the specific problem at hand,
which will be discussed in detail in the following sections.

\item In order to overcome attrition and to reduce the fluctuations 
of $W_n$, one uses ``population control". This is achieved by pruning some 
low-weight configurations and cloning (``enriching"~\cite{Wall-Erpenbeck}) all 
those with high weight, as the chain grows. To define `low' and `high'
weights, one uses two thresholds $W_n^+$ and $W_n^-$. If at any step $n$ 
the current weight $W_n$ according to (\ref{Wn}) would be $>W_n^+$, 
we make $k$ additional copies (typically $k=1$) of the current configuration
and give each copy a weight $W_n = w_n W_{n-1}/(k+1)$. On the other hand, if
(\ref{Wn}) would give $W_n<W_n^-$, we call a random number $r \in \left[0,1\right]$. 
If $r<1/2$ we kill the configuration. Otherwise, we keep it and double its weight,
$W_n = 2 w_n W_{n-1}$. It is easy to see that pruning and cloning leave all 
averages unchanged. It improves importance sampling enormously, but it also 
leads to correlated trials.

For most problems the choice of the thresholds $W_n^+$ and $W_n^-$ is unproblematic, 
and they can be chosen simply as constant multiples of the the current 
estimate of the partition sum given by (\ref{ZM}),
\be 
        W_n^+ = C_+ \hat{Z}_n \qquad {\rm and} 
\qquad  W_n^- = C_-\hat{Z_n} \;, 
\label{eq-Wnt}
\ee
were $C_+$ and $C_-$ are constants of order unity. A good choice for the ratio
between $C_+$ and $C_-$ is found to be $C_+/C_- \sim 10$ in most cases. If 
(\ref{eq-Wnt}) does not lead to good results, chances are that the 
method would not work with any other choice either. If the method works
well, (\ref{eq-Wnt}) gives samples where the total number of length $n$ configurations
is independent of $n$, i.e. attrition is completely eliminated and pruning \& cloning
compensate each other exactly (up to statistical fluctuations), for large $n$.
For the first trials (when there is not yet any estimate $\hat{Z}_n$), we choose 
normally $W_n^-=0$ and $W_n^+=\infty$ (a very large number like $10^{100}$),
which gives the original RR method.
 
\item The copies made in the enrichments are placed on a stack, and a depth-first 
implementation is used for the chain growth: 
At each time one deals with only a single configuration until a chain has 
either grown to the maximum length $N$ or has been killed due to attrition. 
If the first happens or if the stack is empty, a new trial is started. Otherwise, 
the configuration on top of the stack is popped and the simulation continues. This 
is most easily implemented by recursive function calls. Since only a single 
configuration has to be remembered during the run, this requires much less memory
than a breadth-first implementation that uses an explicit ``population" of many 
configurations, as it is traditionally used e.g. in genetic algorithms.

\begin{figure}
\begin{center}
(a)\includegraphics[angle=270,width=0.35\textwidth]{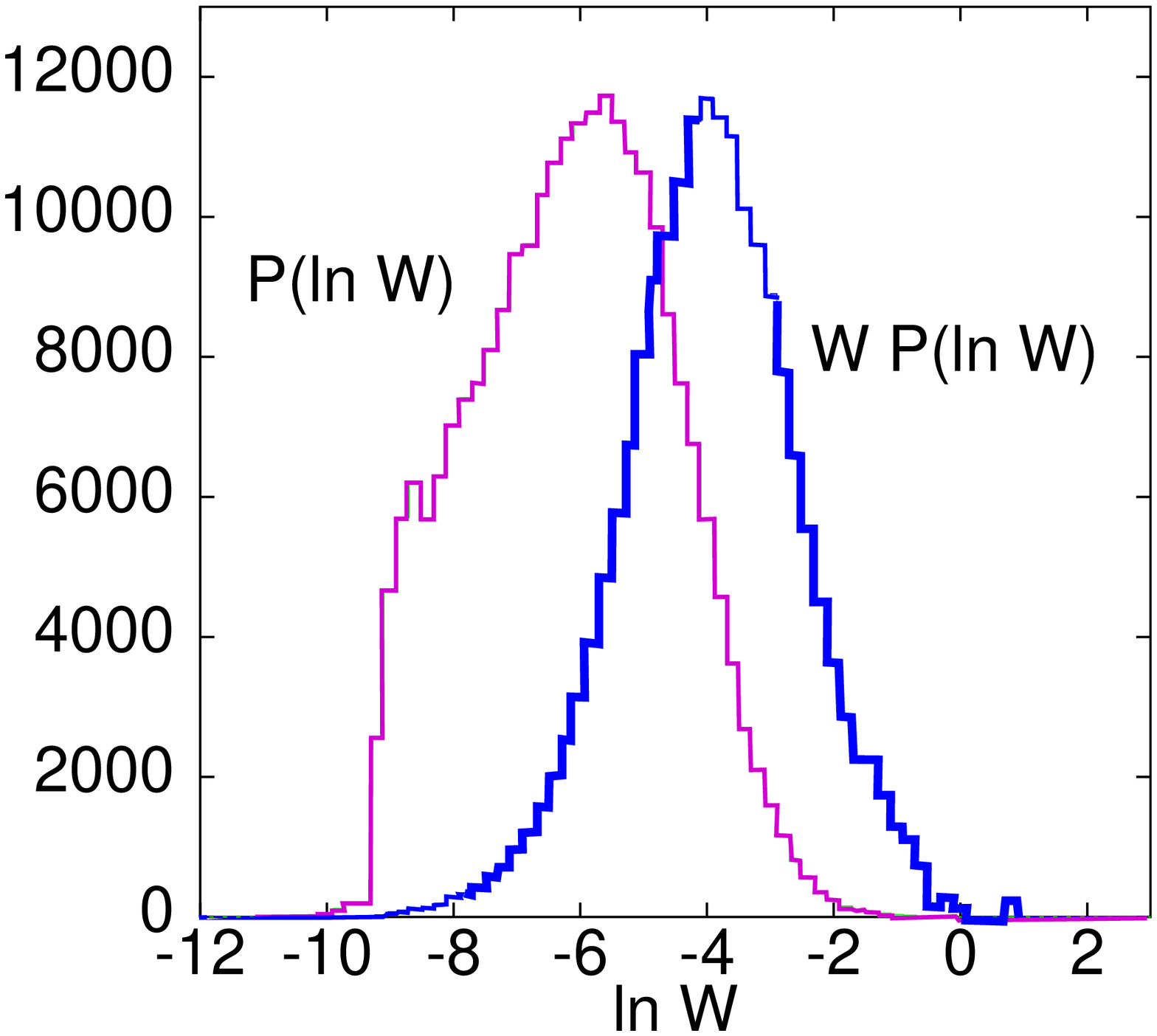}\hspace{1.6cm}
(b)\includegraphics[angle=270,width=0.35\textwidth]{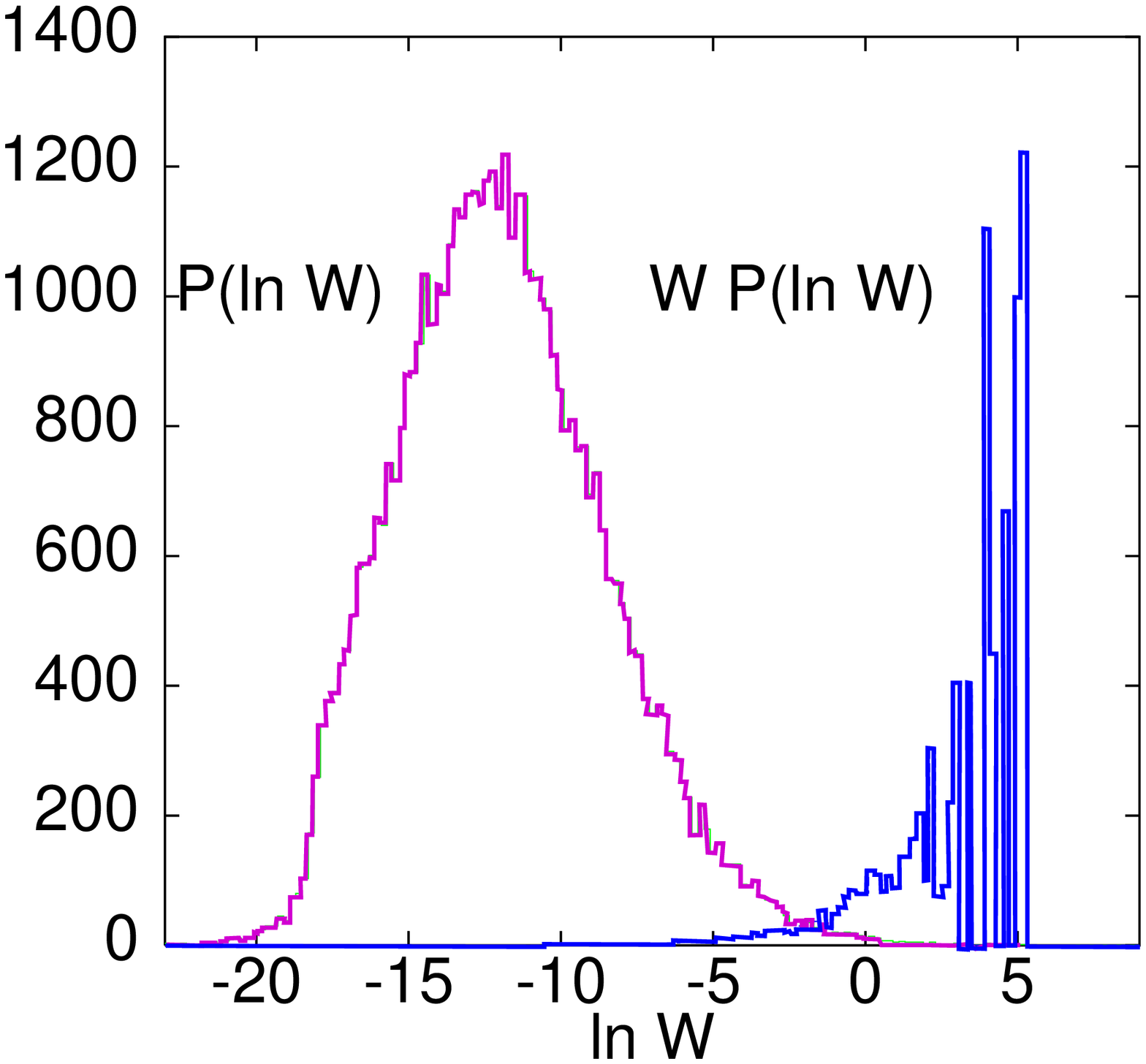}
\caption{Histograms of logarithms of tour weights $P(\ln W)$
normalized as tours per bin, 
and weighted histograms $WP(\ln W)$ are shown as indicated.
Weights $W$ are only fixed up to a $\beta$-dependent
multiplicative constant. The simulation shown in panel 
(a) is reliable, while that in panel (b) is not.
Adapted from Ref.~\cite{PW-G99}.}
\label{fig-pw}
\end{center}
\end{figure}

\item As we said, configurations obtained from different clones of the same 
ancestor will not be uncorrelated. The set of all such configurations is called 
a ``tour". Different tours are uncorrelated. Depending on the amount of 
cloning/pruning, however, the correlations within a tour could be so 
strong as to render the method obsolete. In that case the distributions 
$P(\ln({\cal W}))$ of logarithms of {\it tour weights} ${\cal W}$ is very broad, so 
that we are basically back to the problem of the RR method (with single
trials replaced by tours):  Averages might be dominated, in extreme cases,
by a single tour. For checking against this, we simply look at $P(\ln({\cal W}))$ 
(see Fig.~\ref{fig-pw}), and compare it with the 
weighted distribution ${\cal W} P(\ln {\cal W})$.  
If ${\cal W} P(\ln {\cal W})$ has its maximum at
a value of $\ln {\cal W}$ where the distribution $P(\ln {\cal W})$ 
is well sampled,
we are on the safe side. If not, then the results can still
be correct, but we have no guarantee for it. An illustration of these two 
cases is shown in Fig.~\ref{fig-pw}~\cite{PW-G99}. Fig~\ref{fig-pw}(a)
shows that the sampling is sufficient and the statistical weight distribution
is reliable, but Fig.~\ref{fig-pw}(b) shows the opposite
situation where the result might be completely wrong.

\end{itemize}

\section{Biased Growth}

An important aspect of the method is that in general, for high efficiency, one 
should choose judiciously a bias in the growth, in order to reduce as much as 
possible the fluctuations of the weight factors $w_n$. The optimal choice of bias 
is often a result of trial and error, as there exists no general theory for it.
The two choices discussed in the following subsections are often useful, but 
by no means in all cases -- and other choices may be useful in other applications. 

One aspect of PERM that often decides the success or failure is that any bias 
that improves the growth at an intermediate stage should also be helpful later, 
i.e. it should not lead the growth into a dead end. One application where this 
is violated dramatically is e.g. the problem of random walkers in a medium with
randomly placed traps (the ``Wiener sausage" problem, leading to the famous 
Donsker-Varadhan stretched exponential survival probability~\cite{Donsker}). 
In this problem walkers should, to maximize their survival chance at very long times,
stay very close to their starting point. On the other hand, for short times the 
path integral (partition sum) is dominated by walkers who venture out to explore 
a larger area, even if that might mean they get killed by a trap. Since this system
can be mapped onto a polymer problem, one can apply PERM to it~\cite{Mehra}. 
These PERM simulations gave indeed the first unambiguous numerical verification of the 
Donsker-Varadhan law, nevertheless they completely failed for {\it very} long times, 
because both bias and population control conspired to ``mislead" the 
walkers~\cite{Mehra} to venture too far out.

\subsection{Global Directional Bias}

Assume you want to simulate a polymer whose one end is held fixed at ${\bf x}=0$, and
the other end is pulled away by a constant force ${\bf F}$. In Sect.~\ref{sec-stretch}
we shall discuss in detail the case of a poor solvent where the stretching
might unfold the dense globule into which the unstretched polymer would 
collapse. Here we just discuss qualitatively a polymer in a good solvent,
i.e. a stretched SAW. 

This system could of course be simulated by an unbiased SAW, and the 
stretching could be taken into account by reweighing each obtained 
configurations with a Boltzmann weight $\propto \exp(-\beta{\bf x\cdot F})$.
But this would be extremely inefficient for large $F$, since weights
would be very uneven, and ``correct" configurations would be very rare 
and would have very high weight. 

A much better strategy is to use a bias in the direction of the next step
of the walk in the direction of ${\bf F}$. The amount of the optimal bias
cannot be determined a priori, but depends also on the excluded volume 
effect which helps to push the end further away in the direction of the 
bias. We do not show any data here, but we just mention that the simulations
get easier with increasing $F$, since the walk resembles more and more 
an ordinary biased walk in this limit, and pruning/cloning events get more 
and more rare.

\subsection{PERM with $k$-step Markovian anticipation}
\label{sec-Markovian}
A less trivial bias is suggested by the fact that a growing polymer will 
predominantly grow {\it away} from the already existing part of the chain.
This could be modeled crudely by determining the center of mass of that
part, and biasing the growth away from it. A better strategy is to 
learn on the fly how a typical short existing chain (of $k$ monomers, say) 
would bias the further growth in detail, and to remember at any time 
the previous $k$ steps. This is called 
{\it Markovian anticipation}~\cite{Grassberger98,Frauenkron99,Strip,Slab}.

The crucial point of the $k$-step Markovian anticipation
is that the $(k+1)^{\rm th}$ step of walk is biased by the 
history of the previous $k$ steps, i.e., the bias depends
on the last $k$ steps. Let's consider the general case
of a walk on a $d$-dimensional hypercubic lattice.
At each step $i$, a walk can move towards to one of the $2d$ directions
denoted by $s_i=0$, $\ldots$, $2d-1$.
All possible configurations of $(k+1)$ steps 
($i=-k$, $-k+1$, $\ldots$, $-1$, and $0$), which are in total
$(2d)^{k+1}$ configurations, are labelled by
\be
 \bf S=(s_{-k},\; \ldots,\; s_{-1},\; s_{0})=({\bf s},\; s_{0}) 
\ee
here ${\bf s}$ and $s_{0}$ denote the configurations of 
the previous $k$ steps and the $(k+1)^{\rm th}$ step,
Either during a separate auxiliary run or during the first part of 
a long run we build a histogram $H_m({\bf S})$ with $(2d)^{k+1}$
entries. For any ${\bf S}$, the value of $H_m({\bf S})$ is the 
sum of all contributions to ${\hat Z}_{n+m}$ of configurations that 
had coincided with ${\bf S}$ during the steps $n-k$, $n-k+1$, 
$\ldots$, and $n$, summed over all $n$ in some suitable range excluding 
transients. Typical values for 3-d SAWs might be $k=10, m=100, n>300$.
Then $H_m({\bf S})/H_0({\bf S})$ measures how successful configurations 
ending with ${\bf S}$ were in contributing to the partition sum $m$ steps 
later. The bias in $k$-step Markovian anticipation for the
next step is thus defined by
\be
  P(s_0|{\bf s})=\frac{H_m({\bf s},s_0)/H_0({\bf s},s_0)}
  {\sum_{s_0'=0}^{2d-1} H_m({\bf s},s_0')/H_0({\bf s},s_0')} \;.
\ee

\section{$\Theta$-Polymers}

The first application of PERM was to $\Theta$-polymers in three dimensions
\cite{Grassberger97}. 
As we said, the upper critical dimension for the $\Theta$ collapse is $d=3$,
whence we expect ordinary random walk behavior with logarithmic corrections.
These corrections have been calculated to leading~\cite{Duplantier-Theta} and 
next-to-leading~\cite{Hager} orders. The experimental verification of these
corrections is highly non-trivial, because one has to use extremely diluted
solutions in order to avoid coagulation of different chains, and thus the 
signals are very weak. Nevertheless, they have been observed in small-angle
neutron scattering~\cite{Boothroyd}.

\subsection{A Single $\Theta$-Polymer}

It is for this problem that PERM shows the biggest improvement over all 
other Monte Carlo methods. The reason is that at the $\Theta$ point entropic 
and energetic (Boltzmann-) effects cancel exactly in the limit $N\to\infty$. 
For finite $N$ they do not cancel exactly (this gives rise to the logarithmic
corrections), but it is still true that the weight factors $w_n$ are very 
close to 1. Thus hardly any pruning/cloning is needed, and to a first 
approximation the simulation reduces simply to a straightforward simulation
of random walks with small weight corrections. Full PERM simulations
for very long chains (the longest chains in~\cite{Grassberger97} had $N=10^6$)
do require in average one pruning/cloning step for every 2000 ordinary 
random walk steps. Therefore, in chain length $n$ the algorithm effectively 
performs a random walk with diffusion coefficient $D\approx 2000$. Asymptotically
for $N\to\infty$ the algorithm still needs ${\cal O}(N^2)$ steps to create one 
independent configuration of full length, but the coefficient is tiny.

\begin{figure}
\begin{center}
(a)\includegraphics[angle=270,width=0.45\textwidth]{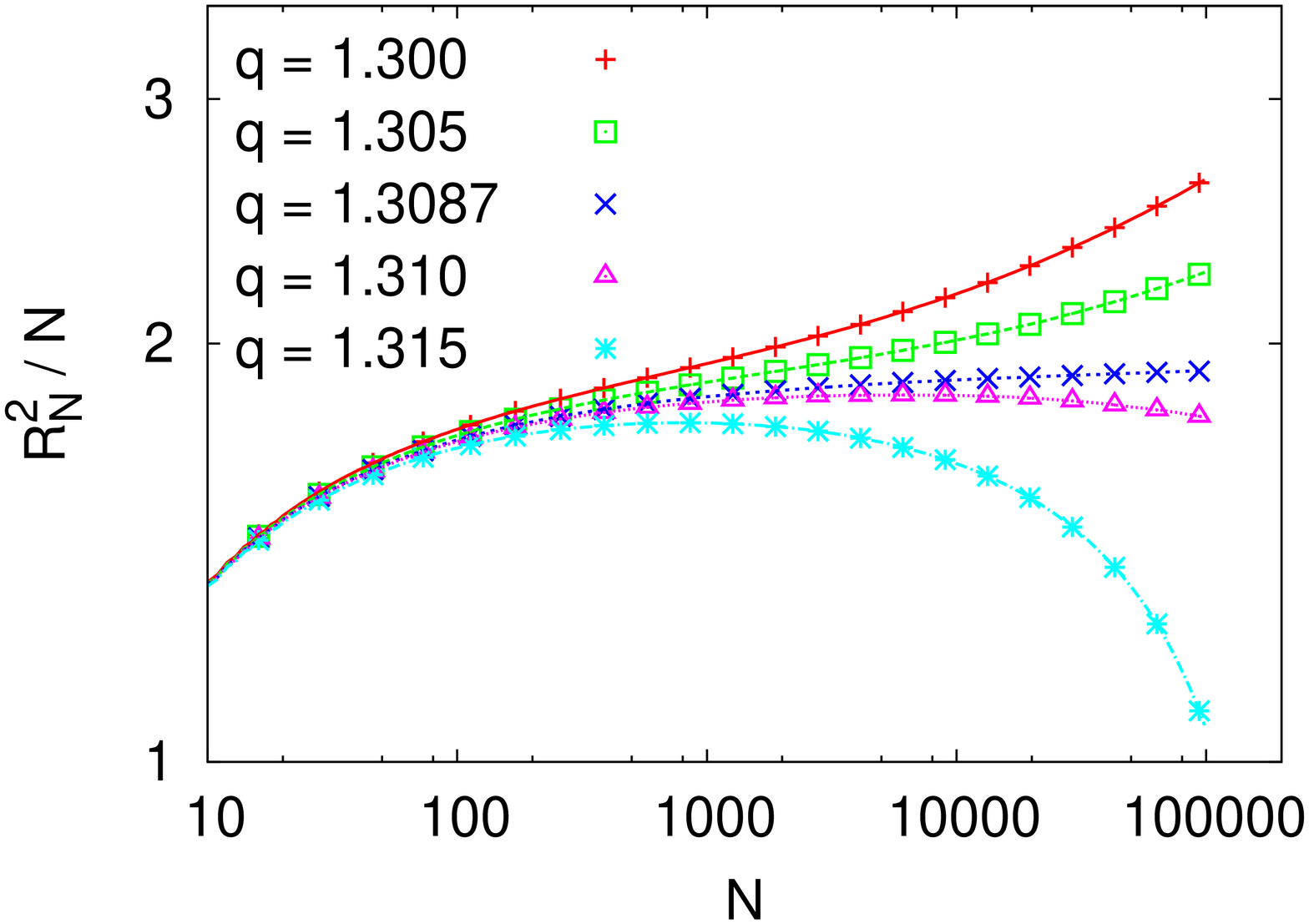}\hspace{0.4cm}
(b)\includegraphics[angle=270,width=0.45\textwidth]{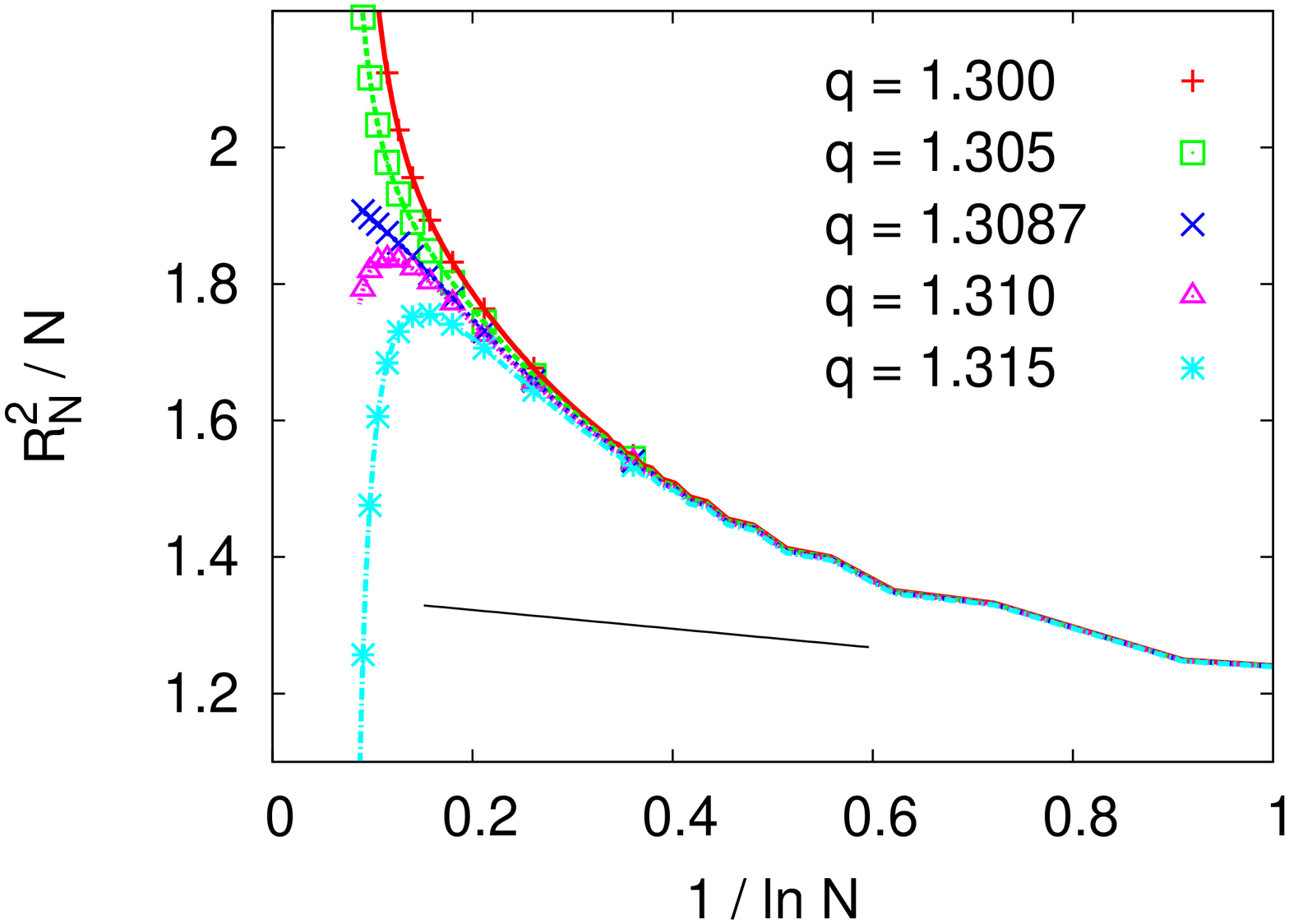}
\caption{The mean square end-to-end distance $R_N^2$ plotted against
$N$ in a log-log scale (a) and against $1/\ln N$ in the normal scale 
(b)~\cite{Grassberger97}. $R^2/N \propto 1-37/(363\ln N)$ is indicated by
the straight line. Adapted from Ref.~\cite{Grassberger97}.}
\label{fig-theta}
\end{center}
\end{figure}

Indeed, since a growing polymer with endpoint in a locally denser region might feel
an elevated Boltzmann factor at step $n$, but feels the compensating entropic disadvantage
only one step later, the optimal algorithm that produced these results was 
a slight modification of the algorithm described in the previous section, where 
the population control was based on a modified weight with incremental weight 
factors 
\be
   w'_n=q^{m_n}/p_{n+1}.     
\ee
instead of (\ref{wn}). Results are shown in Fig.~\ref{fig-theta}.
Theory~\cite{Duplantier-Theta} predicts leading logarithmic corrections to be 
$R_N^2/N \propto 1-37/(363 \ln N)$, which would be a straight line in 
Fig.~\ref{fig-theta}(b) with very small negative slope. Compared to that, the 
corrections to random walk behavior seen in Fig.~\ref{fig-theta}(b) are much larger,
although they are clearly smaller than one would expect for, say, a power law 
correction. It was indeed shown in~\cite{Hager} that the next-to-leading term
increases the deviation from mean field behavior and improves thus the agreement
between theory and simulation, but a fully quantitative verification remains
elusive. 

Far below the $T_\Theta$, PERM becomes inefficient, and it is instructive to see 
why: In strongly collapsed globules, polymer configurations are locally similar to 
those in a dense melt, and are well approximated by simple random walks without
any correlations \cite{deGennes}. But this implies that a collapsed chain
with $N=1000$ has a configuration that is completely different from the first
$1000$ monomers of, say, a collapsed chain of 8000 monomers. The former would 
form a compact globule, while the latter would form a rather dilute structure.
Thus, similar to the problem discussed at the end of the last section, bias and 
population control during the early stages of growth would be completely misleading
as far as late stages of growth are concerned. Otherwise said, by effectively 
disallowing configurations that are initially dilute and fill the interior
only during 
the later growth, the entropy is severely underestimated.

\subsection{Unmixing Transition of Semidilute Solutions of very long Polymers}

Let us now consider a semidilute solution of polymers of common length $N$
slightly below the $T_\Theta$ temperature.
The ``unmixing" transition at which these polymers coagulate and phase 
separate from the solute is, for any finite chain length $N$, in the Ising 
universality class \cite{Widom}. As $N\to\infty$, the transition temperature $T_c$
should approach $T_\Theta$ from below. Since the Ising model has upper critical 
dimension $d_c=4$, but the $\Theta$-point has upper critical $d_c=3$, all critical exponents
referring to collective properties (correlation length, specific heat) should be 
that of the Ising model, while properties characterizing the $N$-dependence (e.g. 
radii of gyration, critical concentration, $T_\Theta - T_c$) should 
be mean field like with logarithmic corrections. In particular, the monomer density
at the critical point should scale as 
\be
    \Phi_c \sim N^{-1/2},                     \label{phi-unmix}
\ee
up to logarithms of $N$.

A long standing problem in the 1990's was that all experiments showed 
$\Phi_c \sim N^{-x_c}$ with $x_c= 0.38\pm 0.01$ \cite{Widom}, which was considered 
as incompatible with theory -- in particular, since experimenters viewed any 
prediction of logarithmic corrections with great skepticism. 

PERM can be easily modified for multi-chain systems, simply by placing the first
monomer of a new chain not near the end of the last chain, and by applying the
correct combinatorial factors that take into account the identities of different 
chains \cite{Frauenkron-unmix}. Such simulations are very inefficient for short
chains, since then $T_c \ll T_\Theta$, but they become more and more efficient 
as $N\to\infty$. They showed clearly that the deviations from (\ref{phi-unmix})
are not due to a different critical exponent, as was believed at this time, but
due to logarithmic corrections \cite{Frauenkron-unmix}. These are much larger than 
predicted by theory \cite{Duplantier92}, but this was to be expected in view of the 
results for single isolated chains.

\section{Stretching Collapsed Polymers in a Poor Solvent}
    \label{sec-stretch}

   As a collapsed polymer chain of chain length $N$ is stretched 
by an external force under poor solvent conditions, one observes 
from a collapsed globule phase to a stretched phase, as the stretching
force is increased beyond a critical value~\cite{Stretch}.
This phase transition is first order in $d=3$ dimensions, as is also 
suggested by the analogy of the Rayleigh instability of a falling 
stream of fluid, but it seems to be second order in $d=2$~\cite{Stretch}.
Here we shall only discuss the 3-d case.

This is modeled as a biased interacting 
self-avoiding random walk (BISAW) on a simple cubic lattice in 
three dimensions. Assuming that a chain is stretched in the x-direction
by the stretching force ${\bf F}=F\hat{e}_x$ ($\hat{e}_x$ is the unit
vector in the x-direction), an additional bias term $b^x$ is 
incorporated into the partition sum given by (\ref{eq-ZNSAW}), where
$b=\exp(\beta a F)$ is the stretching factor ($a$ is the lattice constant)
and $x$ is the distance (in units of lattice constants)
between the two end points of the chain in the direction of ${\bf F}$.
The partition sum is therefore
\be 
     Z_N(q,b) = \sum_{walks} q^m b^x  \;.
\label{eq-st}
\ee
The poor solvent condition is indicated by $q>q_\Theta$ where 
$q_\Theta=e^{-\epsilon/kT_\Theta}\approx 1.3087(3)$~\cite{Grassberger97}. 
According to the scaling law (\ref{Z-}), in the thermodynamic limit 
$N \rightarrow \infty$, the partition sum for polymers in
a poor solvent scales as
\be
     - \ln Z_N(q,b=1) \approx \mu_{\infty}(q)N+\tilde{\sigma}(q)N^{2/3} - (\gamma-1)\ln N
\ee
with $\mu_\infty$ being the chemical potential per monomer in
an infinite chain, and $\tilde{\sigma}$ is related to the surface
tension $\sigma$.

Choosing $q=1.5$ which is deep in the collapsed region,
we performed simulations of BISAW with PERM. 
In order to improve the efficiency,
each step of a walk is guided to the stretching direction 
with a higher probability. 
The $n^{th}$ step of walk (adding the $(n+1)^{\rm th}$ monomer)
is toward one of the free nearest neighbor sites of 
the $n^{\rm th}$ monomer in the 
parallel, antiparallel, and transverse direction
to ${\bf F}$ with probability:
$p_{+}:p_{-}:p_{\perp}=\sqrt{b}:\sqrt{1/b}:1$.
Thus we have
\be 
     p_{i}=\left\{\begin{array}{ll}
 0  \enspace & {\rm if \; the \; step \; of \; the \; walk \; toward \; to }\\
 & {\rm the \; {\it i}-direction \; is \; forbidden} \\
\frac{p_i^{(0)}}{\sum_{{\rm allowed} \,j} p_j^{(0)}} \enspace & {\rm otherwise}
\end{array} \right . 
\label{eq-st-p}
\ee
The corresponding weight factor at the $n^{th}$ step
is then
\be
         w_{i_n}=\frac{q^{m_n}b^{\Delta x_i}}{p_i} \;,
\ee
where $m_n$ is the number of non-bonded nearest neighboring pairs
of the $(n+1)^{\rm th}$ monomer. $\Delta x_i=0$, $1$ or $-1$ if 
the displacement $({\bf r}_{n+1}-{\bf r}_{n})$ between 
the $(n+1)^{\rm th}$ and $n^{\rm th}$ monomers is in the direction
perpendicular, parallel, and antiparallel to ${\bf F}$, respectively.
The total weight of a chain of length $n$ is then
\be
        W_n=\prod_{n'=0}^n w_{i_{n'}}  \;.
\ee 
Using (\ref{ZM}) and (\ref{eq-Wnt}), chains are cloned and
pruned if their weight is above $3\hat{Z}_n$ and below $\hat{Z}_n/3$,
respectively.

\begin{figure}
\begin{center}
(a)\includegraphics[angle=270,width=0.45\textwidth]{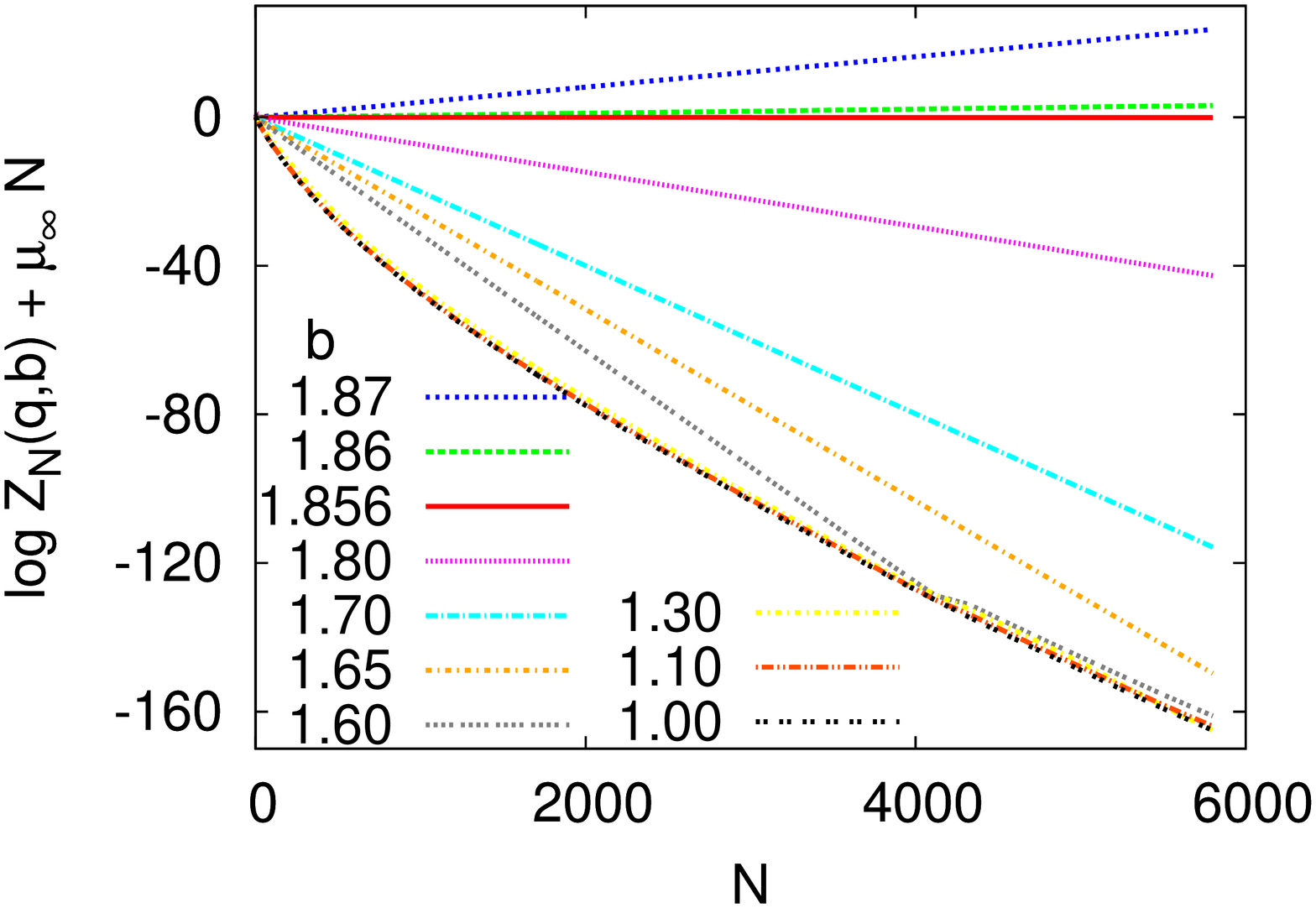}\hspace{0.4cm}
(b)\includegraphics[angle=270,width=0.45\textwidth]{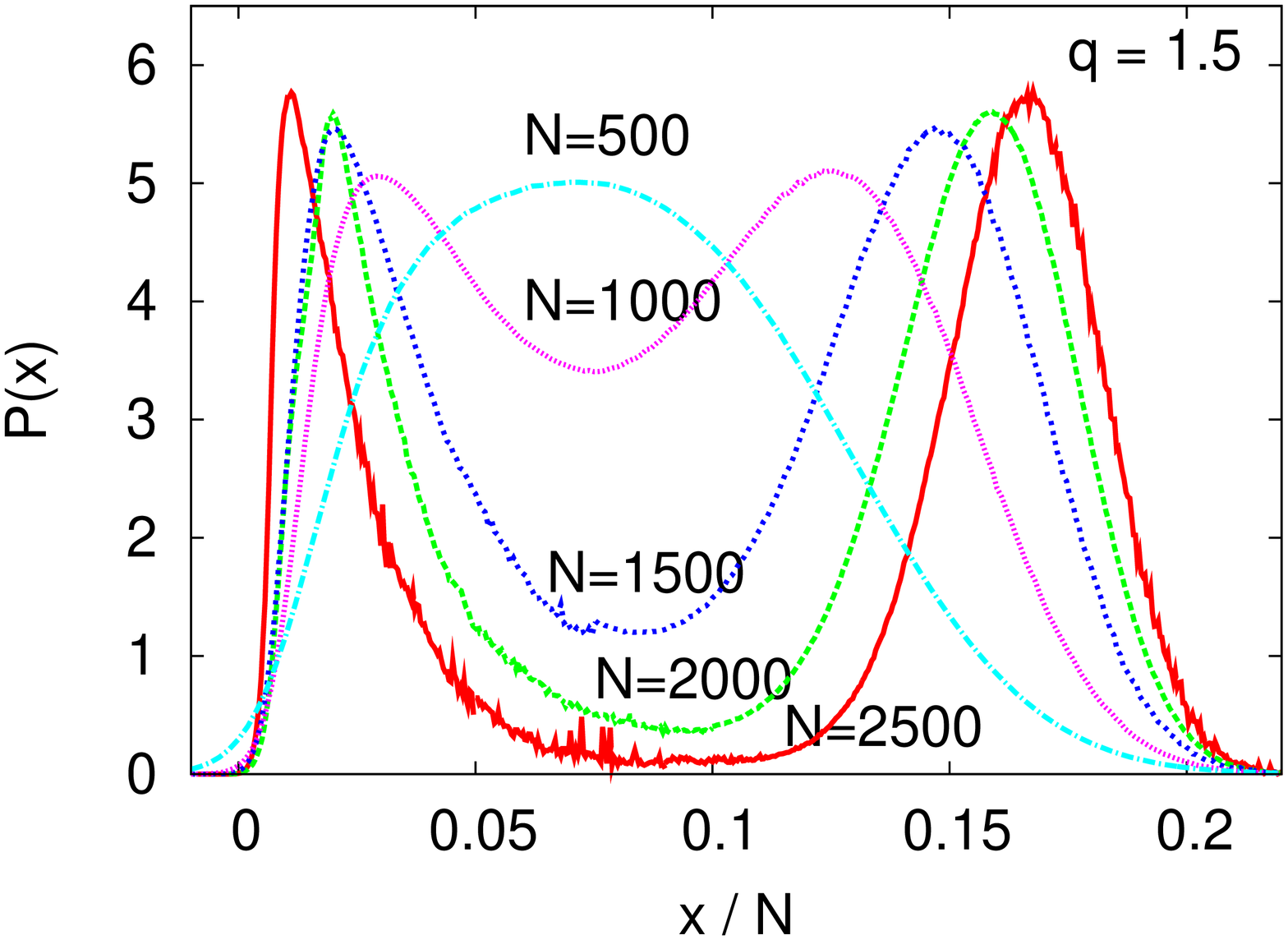}
\caption{(a) $\ln Z_N(q,b) + \mu_\infty N$ for $d=3, q=1.5$, and for various
values of $b$. The value $\mu_\infty=-1.7530\pm 0.0003$ used in this plot
was obtained from dense limit simulations on finite
lattices~\cite{Stretch}.
(b) Histograms of the end point distance $P(x)$ versus $x/N$
for $q=1.5$. Biases were adjusted so that both peaks have
equal height: $b=1.4040\; (N=500)$, $1.4925\; (N=1000)$,
$1.5386\; (N=1500)$, $1.5658\; (N=2000)$, $1.5855\; (N=2500)$.
Normalization is arbitrary. The peak at $x/N \approx 0$ corresponds
to the collapsed phase, the other one to the stretched phase.
Adapted from Ref.~\cite{Stretch}.}
\label{fig-st-zn}
\end{center}
\end{figure}

Results of $\ln Z_N(q,b)+\mu_\infty N$
plotted against $N$ are shown in Fig.~\ref{fig-st-zn}(a)
for various values of $b$.
For small $b$ the curves are close to the curve for $b=1$.
As $b$ increases, the initial (small-$N$) parts of these curves
are straight lines with less and less negative slopes.
In this regime the polymer is stretched. As long
as these slopes are negative, the straight lines will intersect
the curve for $b=1$ at some finite value of $N$, say $N_c(b)$,
i.e. for the finite system of chain length $N_c(b)$ the corresponding
effective transition point is $b$. 
For $N>N_c(b)$, the values of $\ln Z_N(q,b)+\mu_\infty N$ must deviate
from the straight lines \{ see Refs.~\cite{Mehra} and \cite{Stretch} \}
for the detailed explanations. 
Since the curve for $b=1$ becomes horizontal for $N\to\infty$,
the true phase transition occurs at that value of $b$ for which the
straight line in Fig.~\ref{fig-st-zn}(a) is also horizontal. 
This can be estimated very easily and with high precision, giving 
for $q=1.5$ our final estimate $b_c \approx 1.856(1)$. 

To clarify that the transition is indeed a first-order phase
transition, one can study the histograms of $x$ and $m$
since PERM gives direct estimates of the partition sum and
of the properly normalized histograms. The general formula of
the histogram is
\be
   P_{q,b}(m,x) = \sum_{walks}q^{m'}\, b^{x'}\,\delta_{m,m'}\delta_{x,x'} \;.
\ee
Reweighting histograms obtained with runs performed nominally at $q'$ and $b'$
is trivially done by
\be
   P_{q,b}(m,x) = P_{q',b'}(m,x) (q/q')^m (b/b')^x \;.
\label{pmxbb}
\ee
Combining results from different runs can then be either done by selecting for
each $(m,x)$ just the run which produced the least noisy data 
(which was done here in most cases), or by assuming that the 
statistical weights of different runs are proportional to the number 
of ``tours"~\cite{Grassberger97} which contributed to
$P_{q,b}(m,x)$. Note that for conventional Metropolis-type Monte Carlo
algorithms, it is not trivial to combine MC results from different
temperatures since the absolute normalization is unknown~\cite{ferrenberg}.

    An example of histograms $P(x)$ for fixed $q=1.5$ and $b$, plotted against
$x/N$ are shown in Fig.~\ref{fig-st-zn}(b)
for $N=500$, $1000$, $1500$, $2000$, and $2500$.
The value of $b$ is determined such that the two peaks have the 
same height for each $N$, i.e., $b_c(N)=b$ is the effective 
transition point for the finite system of size $N$. 
In addition the normalization factor is chosen arbitrarily 
to make all peaks having similar height for convenience. 
Using (\ref{pmxbb}), each curve in Fig.~\ref{fig-st-zn}(b) contributed
by the properly reweighting data from different runs for various 
values of $b$. Obviously, with increasing $N$, we see that the 
distance between two peaks increases and the minimum between 
the peaks shrinks to zero.
This gives a strong evidence for the first-order transition.  
Notice that a double peak structure with decreasing minimum alone would
not be a conclusive proof, as shown e.g. by the $\Theta$-point in
dimensions $d\ge 4$~\cite{Prellberg00} and by some non-standard 
percolation models~\cite{achliopt}.

\section{Semiflexible Polymer Chains}

\begin{figure}
\begin{center}
\includegraphics[angle=270,width=0.45\textwidth]{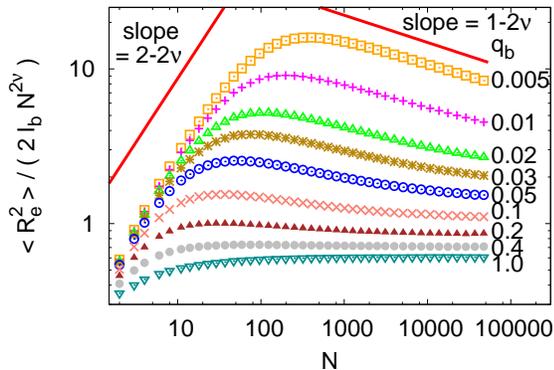}
\caption{Rescaled mean square end-to-end distance
$\langle R_e^2 \rangle / (2 \ell_b N^{2 \nu})$ plotted
against the chain length $N$ for semiflexible chains with
$\ell_b=1$ and various values of $q_b$ on a log-log scale.
Here $\nu \approx 0.588$ is the Flory exponent for SAWs in $d=3$.
Adapted from Ref.~\cite{WLC3d}.}
\label{fig-semiflex}
\end{center}
\end{figure}

Based on a Flory-like treatment~\cite{Flory,Netz}, for a chain with $n$ units   
of the Kuhn length $\ell_K$, and diameter $d$ 
randomly linked together such that the contour length 
$L=N\ell_b=n \ell_K$ (there are $(N+1)$ monomers in the chain
and connected by the bond length $\ell_b$), the effective free energy
of such a semiflexible chain contains two terms as follows,
\be 
   \Delta F \approx R_e^2/(\ell_K L) +v_2R_e^3
\left[(L/\ell_K)/R_e^3\right]^2 \;.
\label{deltaf}
\ee
The first term is the elastic energy which is obtained  
by treating the chain as a free Gaussian chain, hence one 
can immediately write down the probability of the end-to-end distance 
$R_e$ which agrees with the Gaussian
distribution. Therefore, the elastic energy is simply the
logarithm of this distribution.
The second term is the repulsive energy due to 
pairwise contacts where
the second virial coefficient $v_2=\ell_K^2d$,
the density of monomers $\rho=n/R_e^3=LR_e^3/\ell_K$
and the volume $V=R_e^3$.
Minimizing $\Delta F$ with respect to $R_e$, one obtains the 
Flory-type result for self-avoiding walks as $L \rightarrow \infty$
$(N \rightarrow \infty)$
\be
  R_e \approx (v_2/\ell_K)^{1/5} L^{3/5}=(\ell_K d)^{1/5}(N \ell_b)^{3/5} \;.
\ee
The minimum contour length $L$ where the
exclusive volume is effective, i.e. the second term in (\ref{deltaf})
is negligible in comparison with the first one if $N<N^*$,
and using the scaling law of the square for the end-to-end distance 
of a Gaussian chain,
$R_e^2= \ell_KL=\ell_K\ell_b N$,
the upper bound of the chain length for describing the Gaussian chain 
is obtained with
\be
            N^*=\ell_K^3/(\ell_bd^2) \;.
\ee
As $L \leq \ell_K$, the chain shows a rod-like behavior, 
the lower bound of the chain length for the Gaussian chain is given by
$\ell_k/\ell_b$. Therefore, the intermediate Gaussian behavior
should only exist for
\be 
        \ell_k/\ell_b  \leq N \leq N^*  \;.
\label{Gauss}
\ee 
For a linear semiflexible polymer chain ($d=\ell_b$) under  
good solvent conditions,  
one would expect to observe both a crossover from rigid rod-like behavior
to almost Gaussian random coils, then a crossover to
self-avoiding walks when the chain stiffness varies.

In order to verify the prediction, it requires an efficient algorithm 
to generate sufficient samples for very long semiflexible
chains since the results should cover the linear length scales in
the three different regimes. PERM was first applied to this in~\cite{WLC3d}.
The model described below had indeed been studied by means of PERM already 
in~\cite{Theta-WLC}, where however most emphasis was put on the question
whether the collapse transition changes from second to first order 
as the stiffness is increased. This was predicted by mean field 
theories~\cite{Doniach}. The simulations in~\cite{Theta-WLC} supported the 
prediction, but were dangerously close to the significance limit 
discussed in Sect. 2.

The above scaling relations for chains without self attraction 
were studied in~\cite{WLC3d}. Semiflexible 
polymers were there modelled by SAWs on the simple cubic lattice, with
a bending energy $\epsilon_b(1-\cos \theta)$. Here $\theta$
is the angle between the new and the previous bonds (only $\theta=0$
and $\theta=\pm \pi/2$ are possible on a simple cubic lattice).
The partition function of the SAWs of $N$ steps
with $N_{\rm bend}$ local bends (where $\theta=\pm \pi/2$) is
\be
     Z_{N,N_{\rm bend}}(q_b)= \sum_{\rm config.} C(N,N_{\rm bend}) 
q_b^{N_{\rm bend}}
\ee
where $q_b=\exp(−\beta\epsilon_b)$ is the appropriate Boltzmann
factor ($q_b = 1$ for ordinary SAW’s), and $C(N,N_{\rm bend})$
is the total number of chain configurations containing $(N+1)$ monomers
and $N_{\rm bend}$ local bends. 

In the simulation, the walk of length $n-1$ at the
$n^{\rm th}$ step can be guided to either walk straight ahead 
in any direction, or make an $L$-turn. Of course, it is only allowed to 
walk to the free nearest neighbor sites of the $n^{\rm th}$ monomer.
The ratio of probabilities between the 
former case and the latter case is chosen as $1/q_b$.  
Since the stiffness of the chain is controlled by $q_b$, we
give less probability to make an $L$-turn as $q_b$ becomes 
smaller which corresponds to the case that the chain is stiffer.
Results of the rescaled mean square end-to-end distance
$\langle R_e^2 \rangle / (2 \ell_b N^{2 \nu})$ plotted against the chain
length $N$ up to $N=50000$ for $0.005 \leq q_b \leq 1.0$ are 
shown in Fig.~\ref{fig-semiflex}. 
For stiffer chains, namely for smaller values of $q_b$,
we do see a rod-like regime at the beginning for
not very long chains then a cross-over to a Gaussian regime,
and then finally the excluded volume effect becomes more
important for very long chains, and a horizontal plateau
is developed. For very small $q_b$, although the maximum chain
length is up to 50000, it does still not yet reach the SAW regime.
However, this is the first time that one can give
evidence for the existence of the intermediate Gaussian
coil regime (\ref{Gauss}) by using computer simulations.

\section{Polymers in Confining Geometries}

\subsection{Polymers Confined between Two Parallel Hard Walls}
\label{sec-hardwall}

\begin{figure}
\begin{center}
\includegraphics[angle=0,width=0.50\textwidth]{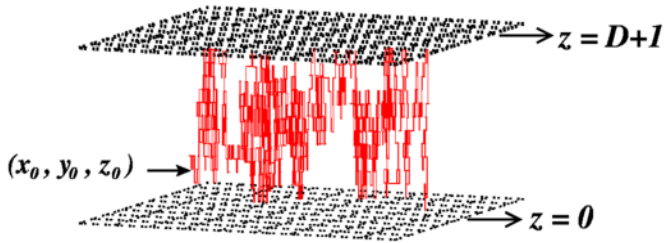}
\caption{Schematic drawing of a polymer chain confined
between two walls located at $z=0$ and $z=D+1$.
For our simulations, chains are grown from the
starting point $(x_0,y_0,z_0)$. Here $x_0$ and $y_0$ are
fixed but $z_0=1,2,\ldots,D$.}
\label{fig-slab}
\end{center}
\end{figure}

  It is a challenge to verify the theoretical scaling predictions for single 
polymer chains of length $N$ confined between two parallel hard walls with 
distance $D$ away from each other (Fig.~\ref{fig-slab}) due to the difficulty 
of producing long 
polymer chains by MC simulations and the existence of very large finite-size 
corrections. For unconstrained SAWs, it is well know that the asymptotic 
scaling behavior is reached rather slowly with correction terms decreasing 
only as $N^{-0.5}$~\cite{LiMadrasSokal,Belohorec,GJPA97}. 
Therefore, in addition 
to SAWs, we studied also the Domb-Joyce (DJ) model~\cite{DJmodel} with 
$v=0.6$ (where convergence to asymptotia is much 
faster~\cite{Belohorec,GJPA97}).

 In the DJ model, polymers are described by lattice walks where monomers sit 
at sites, connected by bonds of length one, and multiple visits to the same 
site are allowed, (i.e. the polymer chain is allowed to cross itself),
but the weight is punished by a repulsive energy $\epsilon>0$ for any pair of 
monomers occupying the same site. Each pair contributes a Boltzmann factor 
$v=\exp(-\beta \epsilon)$ to the partition sum. Thus, the partition sum of 
a linear chain consisting of $N+1$ monomers is given by
\be
     Z_N(v)=\sum_{\rm configs.} v^m \;,
\ee
where the sum extends over all random walk (RW) configurations
with $N$ steps, $0\le v \le 1$, and $m$ is the total number of
monomer pairs occupying a common site,
$m=\sum_{i<j} \delta_{{\bf x}_i {\bf x}_j}$ (${\bf x}_i$ denotes
the position of the monomer $i$). 
For $v=1$, it corresponds to the ordinary RW. For $v=0$ it is
just the SAW model. In the thermodynamic limit where 
$N \rightarrow \infty$, the DJ model
is in the same universality class of SAW for all $v<1$.
There is a ``magic" value of the interaction strength
$v=v^* \cong 0.6$ where corrections to scaling are minimal and
asymptotic scaling is reached fastest~\cite{Belohorec,GJPA97}.
In the renormalization group language, the flow speed
of the effective Hamiltonian approaching its fixed point
depends on $v$. Moreover, it is approached from opposite sides
when $v<v^*$ and when $v>v^*$, with $v^* \cong 0.6$.

There exist important theoretical predictions for the monomer
density profile $\rho(z)$ and the end monomer density profile
$\rho_e(z)$ near the wall given by
Eisenriegler {\it et al.}~\cite{Eisenriegler,Eisenriegler2} as follows:
\be
      \rho(z) \sim z^{1/\nu_3}
\label{eq-wall-rhoz}
\ee
and
\be 
    \rho_{\rm end}(z) \sim z^{(\gamma-\gamma^{(1)})/\nu}
\sim z^{0.814(6)}
\label{eq-wall-rhoendz}
\ee
where $z$ is the distance from the wall and $\gamma^{(1)}$
is the entropic exponent for 3D SAW with one end
grafted on an impenetrable wall.
One should also expect that the density near the walls is
proportional to the force per monomer $f$.
Indeed it was shown by Eisenriegler~\cite{Eisenriegler} that
\be
  \lim_{z \rightarrow 0} k \frac{\rho(z)}{z^{1/\nu_3}}=
  B \frac{f}{k_BT} = B {a\over \nu_3\mu_\infty}\; D^{-1-1/\nu_3}
\label{eq-wall-rhof}
\ee
with $B$ being a universal amplitude ratio. For ideal chains one
has $B=2$, while for chains with excluded volume in $4-\epsilon$
dimensions one has $B \approx 2(1-b_1 \epsilon)$
with $b_1=0.075$~\cite{Eisenriegler3}. In three dimensions this gives
the prediction $B \approx 1.85$.

\begin{figure}[t]
\begin{center}
(a)\includegraphics[angle=270,width=0.45\textwidth]{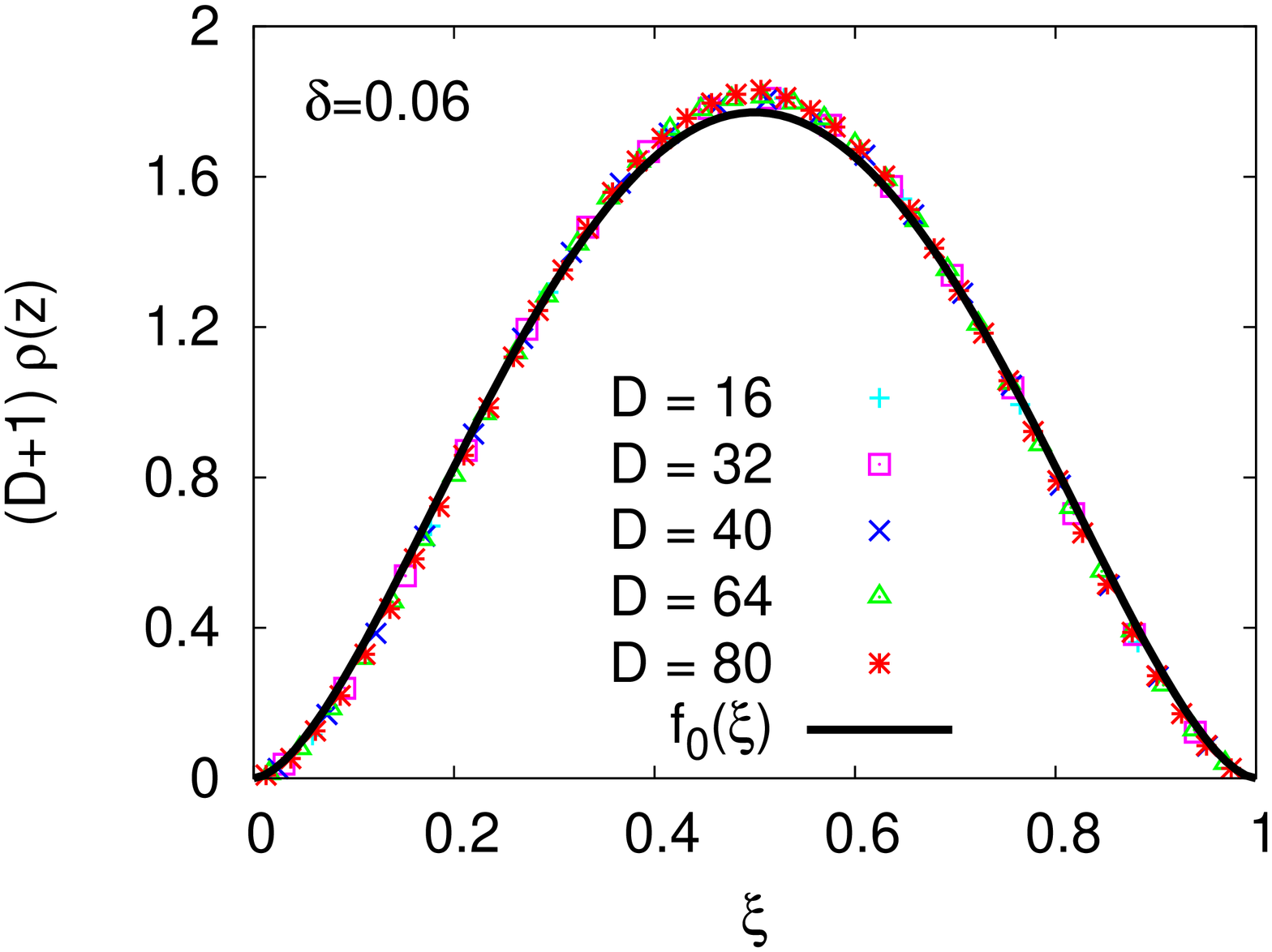}\hspace{0.4cm}
(b)\includegraphics[angle=270,width=0.45\textwidth]{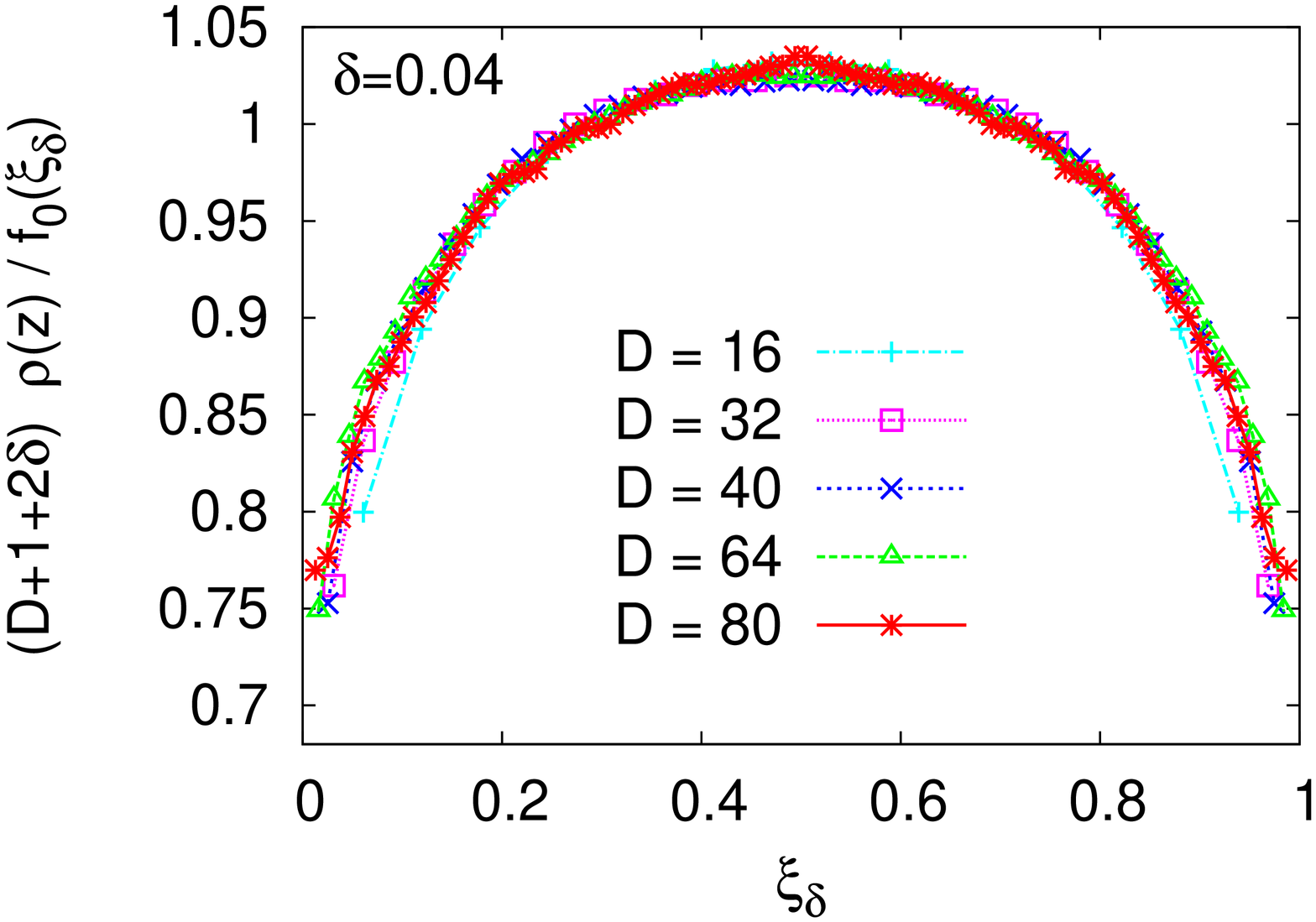}
\caption{Results of the monomer density profiles $\rho(z)$
obtained for the DJ model.
(a) Rescaled values of monomer density
$(D+1)\rho(z)$ plotted against $\xi=z/(D+1)$.
The function $f_0(\xi)=18.74(\xi(1-\xi))^{1/\nu_3}$.
(b) The same data as in (a), divided by $f_0(\xi)$,
plotted against a modified scaling variable,
$\xi_\delta=(z+\delta)/(D+1+2\delta)$ with $\delta=0.04$.
The prefactor in (\ref{eq-wall-rhof})
for $z=0$ and $z=(D+1)\rightarrow \infty$ is
0.71(3). Adapted from Ref.~\cite{Slab}.}
\label{fig-wall-DJ-rho}
\end{center}
\end{figure}

   In order to check the above mentioned theoretical predictions, we
simulate the SAW model and the DJ model on the simple cubic
lattice with the confinement of a slab with width $D$ by
using PERM with $6$-step Markovian anticipation.
For estimating the monomer density profiles $\rho(z)$ we
only count those monomers in the central part of
the chain, excluding $10\%$ on either side to avoid errors
from the fact that (\ref{eq-wall-rhoz}) should hold
only far away from the chain ends, for monomer indices
$n$ satisfying $D^2 \ll n \ll N-D^2$ (we should mention
that $N/D^2>10$ for all data sets).
Results of $\rho(z)$ obtained from the simulations are
normalized such that $\sum_{z=1}^D\rho(z)=1$.
Since we simulate single polymer chains between two walls
at $z=0$ and $z=D+1$, we can assume that
\be
    \rho(z)\approx \frac{1}{D+1} f_0 \left(\frac{z}{D+1}\right)
\enspace {\rm with}
\enspace f_0(\xi)=A\left[\xi(1-\xi)\right]^{1/\nu_3} \;,
\label{eq-wall-rhoz-1}
\ee
where the constant $A=18.74$ is determined by normalization.  We plot
the rescaled values of the monomer density $(D+1)\rho(z)$ against $\xi$ in
Fig.~\ref{fig-wall-DJ-rho}(a) for the DJ model.  It looks like that the scaling
law (\ref{eq-wall-rhoz}) is satisfied and our data are described by
the function $f_0(\xi)$ quite well for $z \in [0,D+1]$.
But, we actually miss the important information near the two walls
in such a plot.  A prefactor on the right hand side of (\ref{eq-wall-rhoz-1})
is probably not a constant. In order to give a precise estimate of the
amplitude $B$ (\ref{eq-wall-rhof}) we introduce here an ``extrapolation
length" $\delta$ as suggested in \cite{Milchev} so that the scaling
variable $\xi$ is replaced by
\be
     \xi_\delta=\frac{z+\delta}{D+1+2\delta} \;.
\ee
Using the same data of $\rho(z)$ but divided by $f_0(\xi_\delta)$, the
best data collapse shown in Fig.~\ref{fig-wall-DJ-rho}(b) is obtained by
taking $\delta=0.04$. 
It leads to
$\lim_{z\rightarrow 0,D \rightarrow \infty} D^{1+1/\nu_3}
z^{-1/\nu_3} \rho(z)/A=0.71(3)$.
Since the extrapolation length $\delta=0.04$ for the DJ model is much smaller
than $\delta=0.15$ for SAWs \{see Fig.~7 in Ref.~\cite{Slab}\}, 
it gives a first indication that corrections to
scaling are indeed smaller in the DJ model. Using (\ref{eq-wall-rhof}),
it gives $B=1.70 \pm 0.08$.  This is only $2$ standard deviations away from
the renormalization group expansion prediction or $\epsilon_c=4-d$ expansion
prediction $B=1.85$ of Eisenriegler~\cite{Eisenriegler},
which we consider as good agreement.

\subsection{Escape Transition of a Polymer Chain from a Nanotube}

\begin{figure*}
\begin{center}
\includegraphics[angle=0,width=0.90\textwidth]{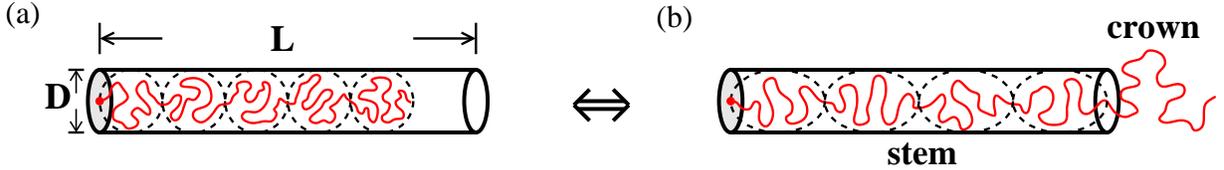}
\caption{Schematic drawings of a flexible polymer chain of length $N$
grafted to the inner wall of a tube of length $L$ and diameter $D$
at the transition point.
(a) As the chain is fully confined in the tube (in an imprisoned
state), it forms a sequence of $n_b=ND^{-1/\nu}$ blobs in a cigar-like shape,
here $\nu=\nu_3$ is the Flory exponent in $d=3$.
(b) As one part of the chain escapes from the tube (in an escaped state),
it forms a flower-like configuration which consists of a ``stem" containing
$N_{\rm tr}$ monomers and a ``crown" containing $N-N_{\rm tr}$ monomers.}
\label{fig-tube}
\end{center}
\end{figure*}

The confinement or escape problem of polymer chains in cylindrical tubes 
of finite length has the merit that it is potentially very relevant to 
experiments and applications such as the problem of polymer translocation
through pores in membranes and the study of DNA confined in artificial 
nanochannels~\cite{Salman,Meller,Kasianowicz}. The following treatment is 
based on \cite{Esc2d,Esc3d-MAC,Esc3d-PRE}.

\begin{figure}
\begin{center}
\includegraphics[angle=270,width=0.45\textwidth]{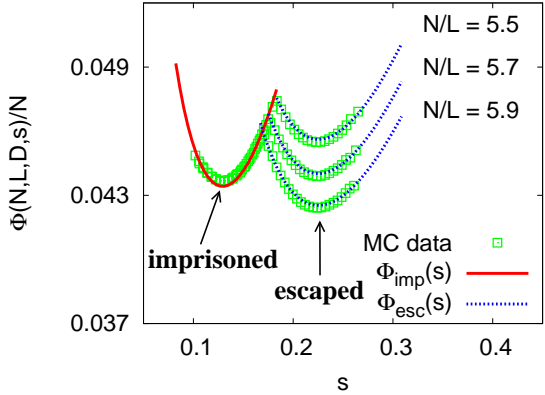}
\caption{The Landau free energy per monomer, $\Phi(N,L,D,s)/N$
plotted against the order parameter $s$ near the transition point
for the tube of length $L=1600$ and diameter $D=17$. }
\label{fig-Landau}
\end{center}
\end{figure}

Considering a polymer chain of length $N$ with one end grafted to the inner 
wall of a cylindrical nanotube with finite length $L$ and diameter $D$ under 
good solvent conditions, the chain configuration is compressed uniformly 
as $D$ decreases or $N$ increases, but $L$ is fixed. Beyond a certain 
compression force, the chain configuration changes abruptly from a 
homogeneously stretched and confined state (imprisoned state) to an 
inhomogeneous state (escaped state) where polymer chains form a flower-like 
configuration with one stem confined in the tube and a coiled crown outside 
the tube (see Fig,~\ref{fig-tube}). This abrupt change implies a first 
order transition. Since the theory based on the blob picture failed to predict
the transition from a homogeneous state to an inhomogeneous state, the 
Landau theory approach is used for describing such a first order transition
including the metastable states.  In the Landau theory approach, all 
configurations are subdivided into subsets associated with a given value of 
an appropriately chosen order parameter $s$ that allows to distinguish 
between different states or phases. The full partition function of the 
system is therefore obtained by integrating over the order parameter:
\be
    Z = \exp(-F)=\int ds \exp [-\Phi(s)] \;,
\ee
where $\Phi(s)$ is the free energy of
a given set, and is therefore a function of the order parameter.
Here the order parameter $s$ is defined by the stretching degree,
i.e. the ratio between the end-to-end distance of monomer
segments which are still confined in the tube, $R_{\rm imp}$,
and the number of monomers confined in the tube, $N_{\rm imp}$.
As shown in 
Fig.~\ref{fig-Landau}, we see that near the transition point the Landau 
free energy function has two minima, the lower minimum is associated with 
the thermodynamically stable state, which corresponds to the equilibrium 
free energy (either $F_{\rm imp}$ or $F_{\rm esc}$) of the system, while 
the other minimum corresponds to the metastable 
state~\cite{Esc3d-MAC,Esc3d-PRE}.
At the transition point, both minima are of equal depth. 

In our simulations, we describe the grafted single polymer chain confined 
in a tube by SAWs of $N$ steps on a simple cubic lattice with cylindrical 
confinement $\{0\leq x \leq L,\, y^2+z^2=D^2/4\}$, and the first monomer is 
attached to the center of the inner wall of the tube. 
Taking the advantage of PERM that the associated weight of each generated 
configuration is exactly known, we introduce a new strategy in order to 
obtain sufficient samplings of the flower-like configurations in the phase 
space as follows: We first apply a constant force along the tube to pull the 
free end of a grafted chain outward to the open end of the tube as long as 
the chain is still confined in a tube, and release the chain once one part 
of monomer segments of it is outside the tube. Varying the strength of the 
force, we obtain flower-like configurations containing stems with various 
stretching degree of monomer segments which are still confined in a tube if
the length $N$ is long enough. The contributions for the escaped states are
therefore given by properly reweighting these configurations to the situation
where no extra force is applied. This is done by using biased SAWs (BSAWs) on 
a simple cubic lattice with finite cylindrical geometry confinement,
similar to the model in (\ref{eq-st}), but we use here 
$q=1$ to describe the good 
solvent condition.

With PERM, the total weight of a 
BSAW of $N$ steps ($N+1$ monomers) is $W_b(N,L,D)=\Pi_{n=0}^N w_n$ with 
$w_n=b^{(x_{n+1}-x_n)}/p_n$ for $n\ge 1$ and $w_0(N,L,D)=1$. $p_n$ is chosen 
as in (\ref{eq-st-p}). The estimate of the partition sum is given by
\be
 \hat{Z}_b(N,L,D)= \frac{1}{M_b}\sum_{{\rm configs.}\in{\cal C}_b} 
W_b({\cal C}_b) \,
\ee
where a set of configurations is denoted by ${\cal C}_b$. Thus, each 
configuration of BSAWs with the stretching factor $b_k$ contributes a weight 
$W^{(k)}(N,L,D)$ for a BSAW of $N$ steps with $b=1$ confined in a finite tube 
of length $L$ and diameter $D$:
\be
        W^{(k)}(N,L,D) = \left\{\begin{array}{ll}
W_{b_k}(N,L,D)/b_k^{x_{N+1}-x_{1}}\;, & x_N \le L \\
W_{b_k}(N,L,D)/b_k^L \;,        & x_N > L
\end{array} \right . \;,
\ee
where index $k$ labels runs with different values of the stretching factor $b$.
Combining data runs with different values of $b$, the final estimate
of the partition sum is 
\be
  Z(N,L,D)=\frac{1}{M} \sum_k \sum_{configs. \in {\cal C}_{b_k}} W^{(k)}(N,L,D)
\label{ZNLD}
\ee
here $M$ is the total number of trial configurations.

The distribution of the order parameter, $P(N,L,D,s) \propto H(N,L,D,s)$,
is obtained by accumulating the histograms $H(N,L,D,s)$ of $s$,
where $H(N,L,D,s)$ is given by,
\be
\begin{array}{lll}
H(N,L,D,s) &=&  \frac{1}{M}\sum_k H^{(k)}(N,L,D,s) \nonumber \\
&=& \frac{1}{M}\sum_k \sum_{configs.} \in {\cal C}_{b_k}
W^{(k)}(N,L,D,s')\delta_{s,s'}
\end{array}
\ee
and the partition sum of polymer chains confined in a
finite tube can be written as
\be
           Z(N,L,D)=\sum_s H(N,L,D,s)
\ee
in accordance with (\ref{ZNLD}). Thus, one can also double check 
the results of the partition sum. 

The Landau free energy $\Phi(N,L,D,s)$ here is the excess 
free energy related to the polymer chains with one end tethered
to an impenetrable flat surface, i.e.
$\phi(N,L,D,s)=-\ln\left[P(N,L,D,s)/Z_1(N)\right]$
($Z_1(N) \sim \mu^{N}N^{\gamma_1-1}$~\cite{Graf-G05}).
Results shown in Fig.~\ref{fig-Landau} are for $L=1600$,
$D=17$, and for $N/L=5.5$, $5.7$, $5.9$.
This shows that the information about metastable states can also be extracted
from the simulations with PERM.
 
\begin{figure*}
\begin{center}
\includegraphics[angle=0,width=0.90\textwidth]{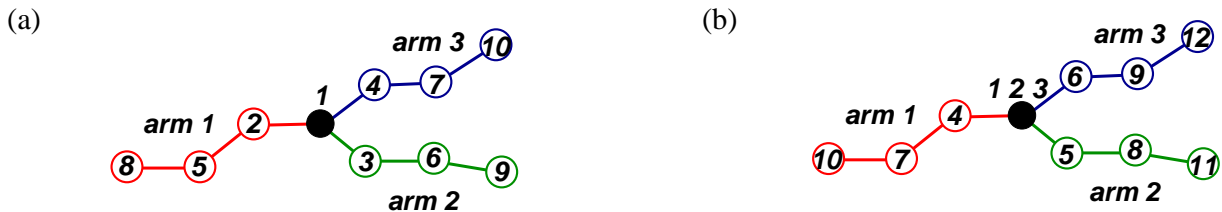}
\caption{Schematic drawings of a star polymer consisting
of three arms $(f=3)$ of length $N=3$ each.
The center is singly occupied in (a) and $f$-folded occupied in (b).
Those numbers show the order of monomers which is added into
the star polymer by using a chain growth algorithm.}
\label{fig-star}
\end{center}
\end{figure*}

\section{PERM for Branched Polymers with Fixed Tree Topologies}
\label{sec-Multiarm}

In this section we shall discuss two types of branched tree-like polymers:
Star polymers (where all branches emanate from one single point) and ``bottle 
brushes" where side chains of common lengths are attached to a backbone at 
regularly spaced points.

To be concrete, let us consider the simplest case of a branched polymer, 
a star polymer where $f$ arms are grafted to a single branch point, 
and all arms have the same length $N$.

As a linear chain is built by using PERM, at each step one monomer is added 
to the built chain until the chain has reached its maximum length $N$ or it 
has been killed in between. For growing a star polymer we have to be aware 
that not only the interactions between monomers in the same arm have to be 
considered but also the interactions between monomers on different arms have
to be taken into account. If one arm is grown entirely before the next arm is 
started, it will lead to a completely ``wrong" direction of generating the 
configurations of a star. However, it is straightforward to modify the basic 
PERM algorithm such that all $f$ arms of a star polymer are grown 
simultaneously~\cite{Star1,Star2}.
The multi-arm method is explained as follows:

\begin{itemize}
\item A star polymer is grown from its branching point (center).

\item $f$ growth sites $\{{\bf x}_1, \ldots {\bf x}_f\}$ are considered 
at the same time. A monomer is added to each arm step by step until all arms 
have the same length, then the next round of monomers is added.
As all the monomers in a star are numbered, it is similar as growing one 
linear chain from the $1^{\rm st}$ monomer to the $N_{\rm max}^{\rm th}$ 
monomer (see Fig.~\ref{fig-star}). $N_{\rm max}=Nf+1$ if the center is singly 
occupied or $N_{\rm max}=Nf+f$ if the center is $f$-folded occupied.

\item A bias is given to guide the growth of arms into outward direction with 
higher probability. The strength of this bias is adjusted in the way that
it increases with $f$ but decreases as the length of arms becomes longer since 
there is more space in a dilute solvent for adding the next monomer. 
For example, we can choose the bias as a function of $n$, 
$g(n)$, for $n \ge 0$,
\be
   g(n)=\left\{\begin{array}{ll}
     (n+4.0)/(n+1.3) \;,\enspace & \enspace {\rm outward \, direction} \\
     (n+0.6)/(n+3.9) \;, \enspace & \enspace {\rm otherwise}
\end{array} \right . \; .
\ee
However, the strength of this bias can be adjusted by trial and error.

\item The population control (pruning/cloning) is done in the same way as 
explained in Sect.~\ref{sec-PERM} that at the step $n$, two thresholds $W_n^+$ 
and $W_n^-$ are proportional to the current estimate weight $\hat{Z}_n$, e.g.,
$W_n^+=3 \hat{Z}_n$ and $W_n^-=0.5\hat{Z}_n$. 

\end{itemize}

\subsection{Star Polymers}
  
   For single star polymers composed of $f$ arms of length $N$ each
in a good solvent, the partition sum and the rms center-to-end
distance scale as follows:

\be
      Z^{(1)}_{N,f} \sim \mu_\infty^{-fN} N^{\gamma_f-1}
\label{eq-star-ZN}
\ee
and
\be
      R^2_{N,f} \approx A_f N^{2 \nu}
\label{eq-star-R2N}
\ee
where the critical fugacity $\mu_\infty$ and the Flory exponent $\nu$ are the 
same for all topologies but the entropic exponent $\gamma_f$ depends on each 
topology~\cite{Duplantier}. 
In two dimensions, $\gamma_f$ can be calculated exactly 
by using conformal invariance~\cite{Duplantier}, but
there are no exact results for the $f$-dependent power law
for $\gamma_f$, and also not for the swelling factor $A_f$.
Therefore, computer simulations are needed for a deep understanding
of star polymers. 
Due to the difficulty of simulating the star polymers with 
many arms $f$ and of long arm length $N$ 
by both MC simulations~\cite{Barrett,Batoulis,Shida,Dicecca,Ohno,Zifferer}
and molecular dynamics~\cite{Grest,Grest94},
and because of the lack of precise estimates of the exponents 
given in ({\ref{eq-star-ZN}) and ({\ref{eq-star-R2N}),
PERM with multi-arm growth method as explained above
was developed~\cite{Star1}. With this algorithm, high
statistics simulations are obtained for star polymer with
arm number up to $f=80$ and arm length up to $N=4000$ for
small values of $f$.

   For our simulations of single star polymers in a good solvent, we use 
the Domb-Joyce model with the interaction strength $v^*=0.6$ 
on the simple cubic 
lattice (see Sect.~\ref{sec-hardwall}). It allows us to attach a larger number 
of arms to a point-like center of stars, and thus additional 
considerations of the corrections to scaling terms when a finite size  
core is used are avoided. Two variants 
for studying star polymers are used in our simulations. In one variant
the center is occupied by one monomer, and in the other variant 
the center is occupied by $f$ monomers as shown in Fig.~\ref{fig-star}. 
Since the partition sum is 
estimated directly by PERM, the exponents $\gamma_f$ can also be determined 
easily according to ({\ref{eq-star-ZN}).  

In Fig.~\ref{fig-star-gamma}, we present
results of $\gamma_f$ from our simulations and from previous 
studies~\cite{Batoulis,Shida,Schaefer92} for comparison. 
The theoretical prediction for the scaling law of $\gamma_f$ for large $f$ 
by the cone approximation~\cite{Ohno,Witten} is 
\be
          \gamma_f-1 \sim f^{-3/2}
\label{eq-star-cone}
\ee
For small $f$, our results are in good agreement
with the previous studies. For large $f$ the best fit with a power 
law $\gamma_f-1 \sim -(f-1.5)^z$ would be obtained with $z \approx 1.68$,
which is not too far off the theoretical prediction (\ref{eq-star-cone})
but the prediction is also not exact.

\begin{figure}
\begin{center}
(a)\includegraphics[angle=270,width=0.45\textwidth]{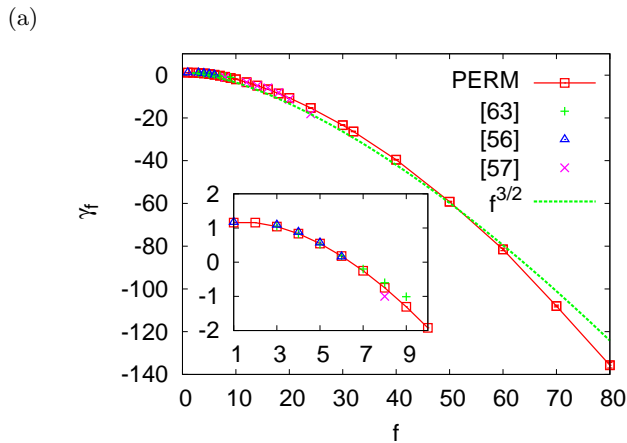}\hspace{3.0cm}
\caption{Exponents $\gamma_f$ plotted against $f$. 
The full line is just a polygon connecting the points, and the dashed line 
is a fit with the large-$f$ behavior as predicted by the cone approximation 
(\ref{eq-star-cone}). Results obtained in Ref.~\cite{Batoulis,Shida,Schaefer92}
are shown for comparison. In the inset, we show those results for small $f$.
Adapted from Ref.~\cite{Star1}.}
\label{fig-star-gamma}
\end{center}
\end{figure}

     After we have obtained quite reliable estimates of $\mu_\infty$, $\nu$,
and $\gamma_f$ for single star polymer in a good solvent, 
we extend our study to a more complicated system where two star polymers 
interact with each other~\cite{Star2} by using the same model and the
same algorithm. It is well understood that interactions between both linear and
branched polymers are soft in the sense that they can penetrate each other 
and the effective potential is a rather smooth function of their distance.
For star polymers, there are some contradictions between results in the 
literatures. Is the potential between two central monomers at large distance  
a Gaussian potential or has it a Yukawa tail? Since we were able to simulate
star polymers up to $f=80$ arms, we expected that we would give a clear answer.
This was the main motivation to study the effective potential between two 
star polymers~\cite{Star2}. 
 
Witten and Pincus~\cite{Witten}
point out that the scaling of the partition sum of a star with $f$ arms
and arm length $N$ each (\ref{eq-star-ZN}), together with the assumption
that $Z^{(2)}_{N,f}(r)/\left[Z_{N,f}^{(1)}\right]^2$ is a function of 
$x\equiv r/R_g$ only for any fixed $f$, i.e.
\be
        \frac{Z^{(2)}_{N,f}(r)}{\left[Z_{N,f}^{(1)}\right]^2}
=\psi_f(r/R_g)  \;,
\ee
implies that 
\be
    V(r) \approx V_{\rm WP}(r) \equiv b_f \ln (a_fR_g/r)
\label{eq-star-V}
\ee
where $r$ is the distance between the two central monomers, and 
\be
        b_f=(2\gamma_f-\gamma_{2f}-1)/\nu  \qquad {\rm for}
\enspace 1 \ll r \ll R_g \;.
\ee
According to our results shown in Fig.~\ref{fig-star-gamma}, instead
of the scaling $b_f \sim f^{3/2}$, a power law gives 
$b_f \approx 0.27 f^{1.58}$. However, both $a_f$ and $b_f$ should be 
universal and should not depend on the specific microscopic realization.

\begin{figure}
\begin{center}
(a)\includegraphics[angle=270,width=0.45\textwidth]{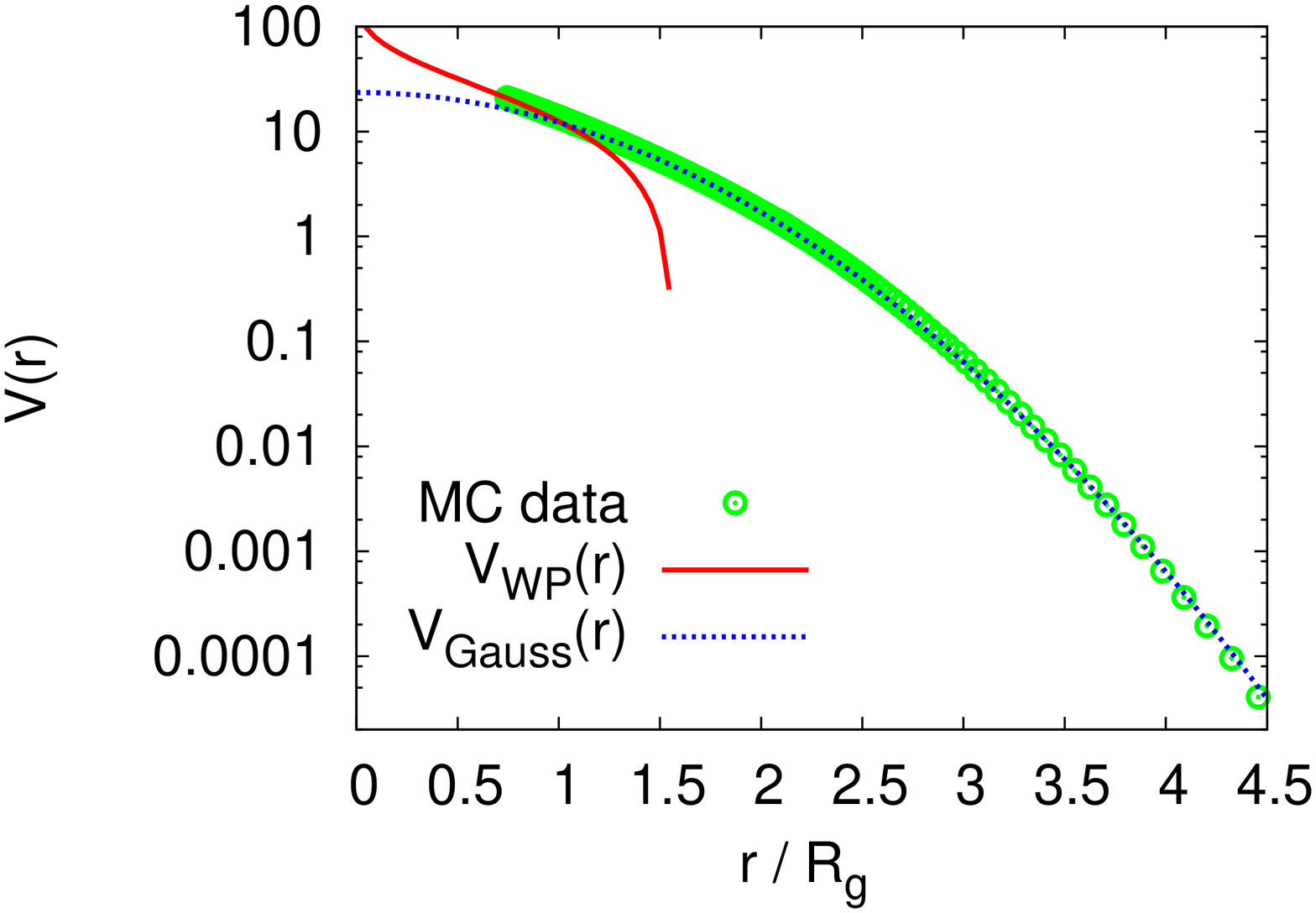}\hspace{0.4cm}
(b)\includegraphics[angle=270,width=0.45\textwidth]{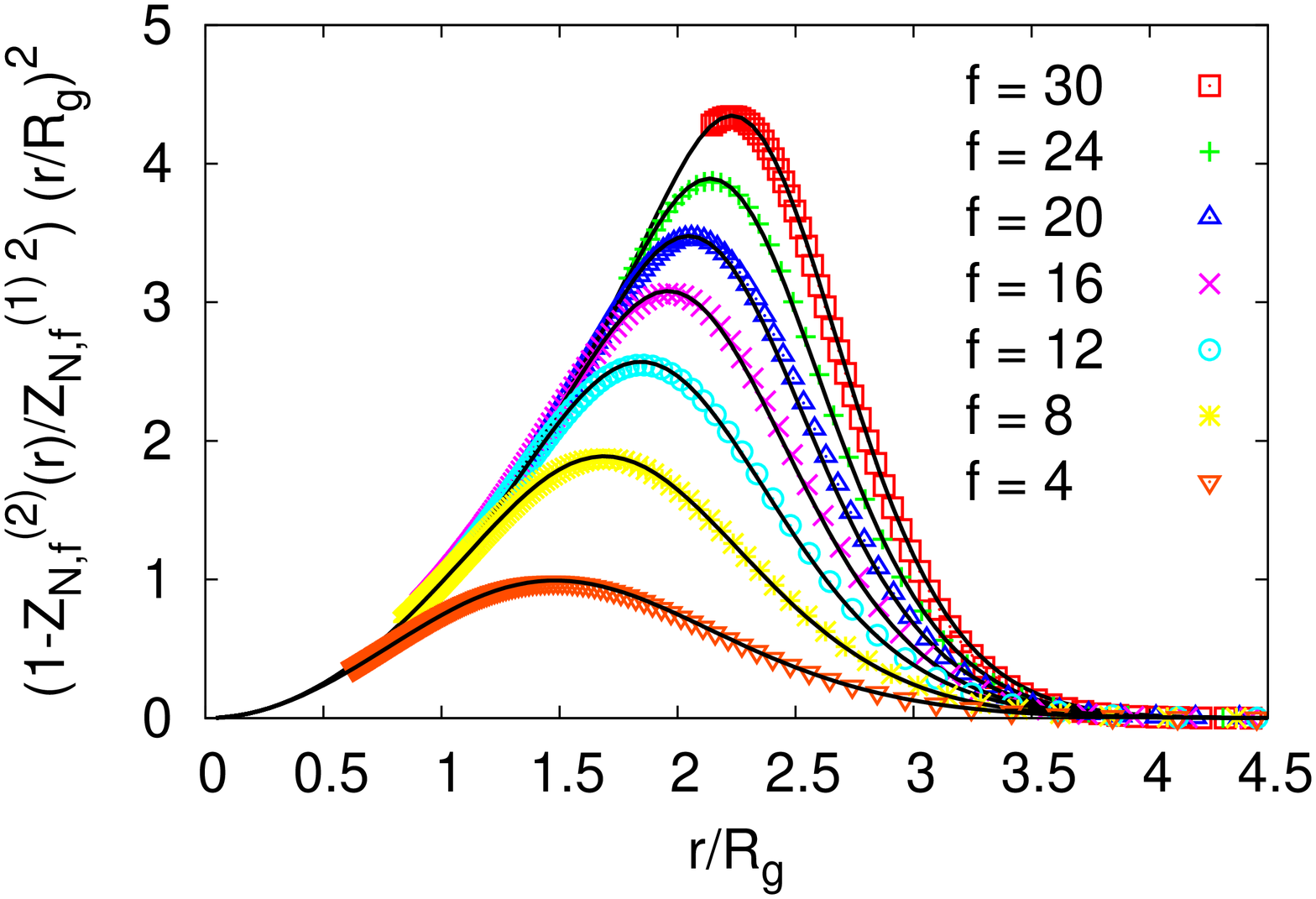}
\caption{(a) The effective potential $V(r)$ for $f=18$, plotted
against $R_g$ in a semi-log scale. The solid curve shows (\ref{eq-star-V}),
and the dotted curve is a Gaussian.
(b) Rescaled radial Mayer functions against $r/R_g$ for several values of $f$.
Curves are obtained from (\ref{eq-star-Mayer}), with fitted parameters
$a_f$, $c_f$, $d_f$ and $\tau_f$. 
Adapted from Ref.~\cite{Star2}}
\label{fig-star-V}
\end{center}
\end{figure}
 
There are two methods for estimating $Z^{(2)}(r)$ in our simulations:
\begin{itemize}
\item[(a)] Two independent star polymers are grown simultaneously, and 
$Z^{(2)}(r)$ is computed by counting their overlaps at different distance $r$. 
Here $Z^{(2)}(r)$ and $Z^{(1)}(r)$ are estimated in the 
same run, which gives rather accurate results for the potential $V(r)$
for very large distances $r$ and large $N$. For small distances $r$,
the ratio $Z^{(2)}(r)/\left[Z^{(1)}(r)\right]^2$ would be indistinguishable 
from zero.
\item[(b)] Two star polymers are grown at fixed distance $r$ with the mutual 
interactions taken into account during the growth.
This allows us to measure $Z^{(2)}(r)$ down to very small distances $r$
and large $N$. For large distances $r$, it gives very bad results of 
the potential $V(r)$ since it is obtained by subtracting the (nearly equal)
free energies obtained in two different runs.
\end{itemize}
For the data analysis, we use the data either from the first method or
the second method, or use the combination from both.

    We present the effective potential $V(r)$ between two
star polymers of $f=18$ arms in Fig.~\ref{fig-star-V}(a).
For $r \ll R_g$, $V(r)$ follows the prediction given in (\ref{eq-star-V}),
which is shown by the solid curve. For $r \gg R_g$ the MC data can be
approximated by a parabola, i.e. $V(r)$ is roughly Gaussian
\be    
       V(r) \approx V_{\rm Gauss}(r) \equiv c_f e^{d_f r^2/R_g^2} \;.
\label{eq-star-VG}
\ee
Here we conjecture that $c_f$ and $d_f$ are universal.
In order to describe the effective potential $V(r)$ for the whole
region of $r$, we propose that 
\be
      V(r)=\frac{1}{\tau_f}\ln \left[ e^{\tau_f V_{\rm WP}(r)-d_fr^2/R_g^2}+
e^{\tau_f V_{\rm Gauss}(r)} \right] \;,
\ee
where $\tau_f$ is an additional parameter for every $f$,
and $V(r) >0$ for all $r$. As $r\rightarrow \infty$, 
$V(r)=V_{\rm Gauss}(r)[1+O(r^{-b_f})]$ (\ref{eq-star-VG}), while 
$V(r)=V_{\rm WP}(r) [1+O(r^2)]$ (\ref{eq-star-V}) as $r \rightarrow 0$.
In fig~\ref{fig-star-V}(b), we plotted the rescaled radial Mayer function,
\be
    (r/R_g^2)^2 f_M(r)=(r/R_g)^2(a-\exp[-V(r)]) \;,
\label{eq-star-Mayer}
\ee
against the rescaled distance $r/R_g$. Our results are in good agreement
with the simulations of~\cite{Rubio} but do not agree with the results
in~\cite{Likos}.

\begin{figure*}
\begin{center}
\includegraphics[angle=0,width=0.90\textwidth]{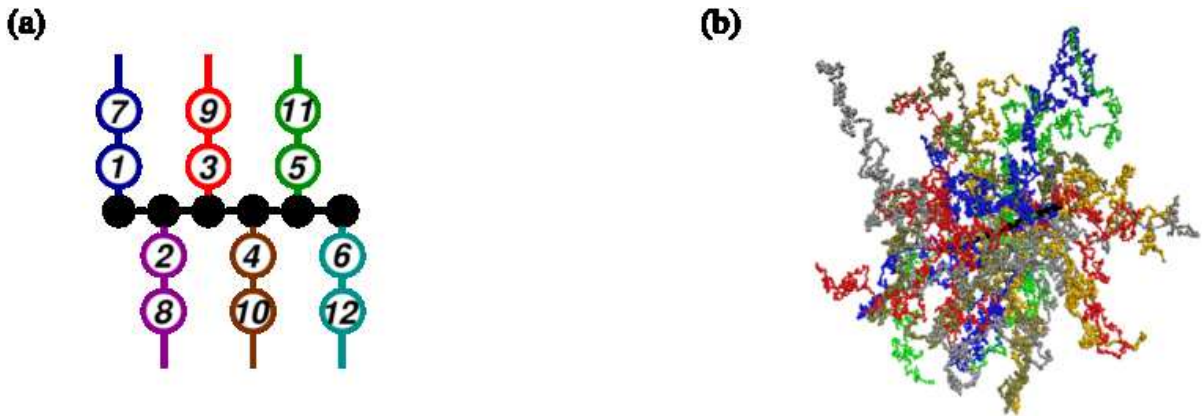}
\caption{(a) A schematic drawing of growing a bottle-brush polymer
step by step.
(b) A snapshot of the configurations of bottle-brush polymers
consisting of $N_b=128$ backbone monomers, $N=2000$ monomers in
each side chain, and the grafting density $\sigma=1/4$ 
under a very good solvent condition generated by PERM.}
\label{fig-bottlebrush}
\end{center}
\end{figure*}

\begin{figure}
\begin{center}
(a)\includegraphics[angle=270,width=0.45\textwidth]{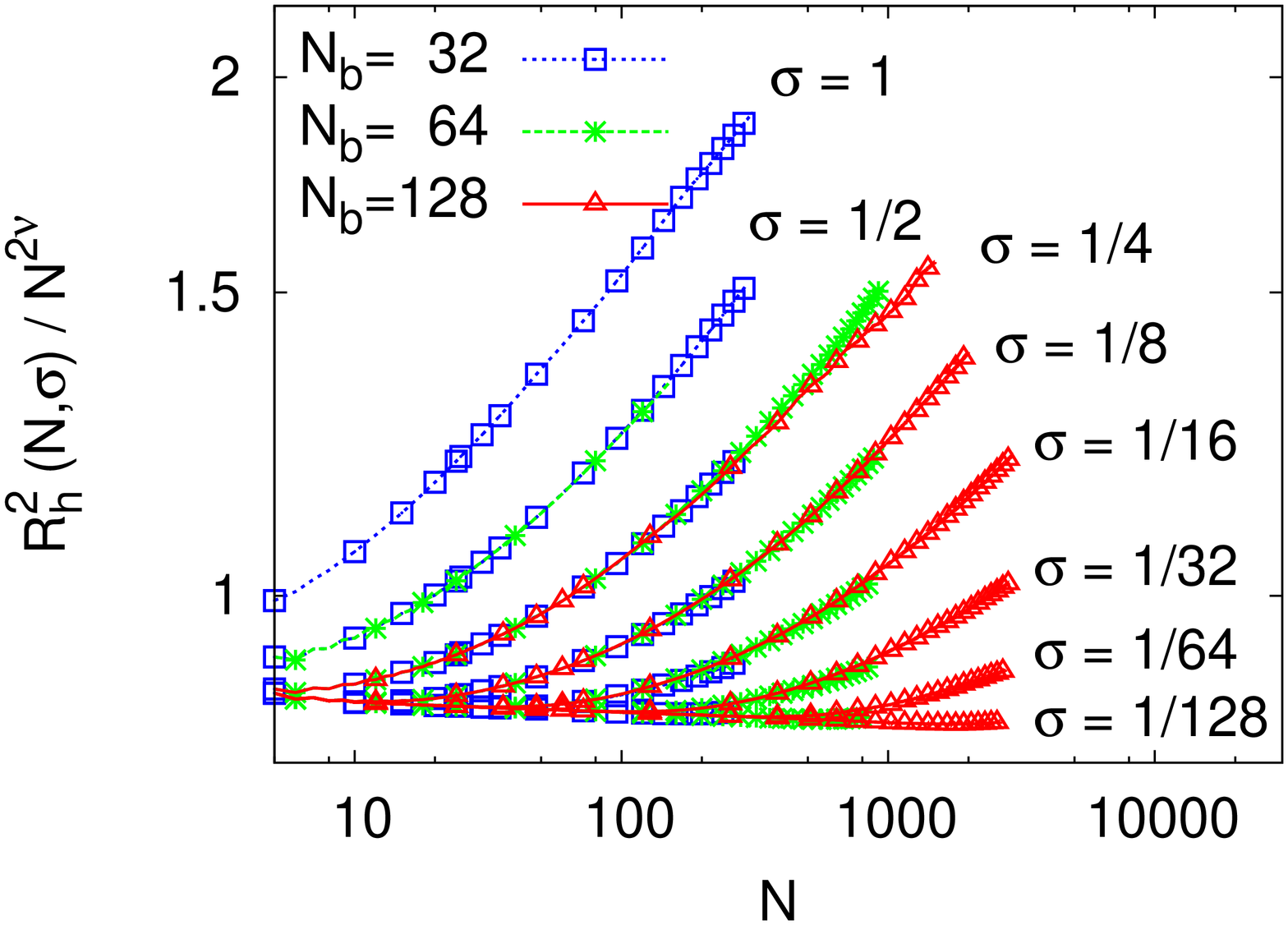}\hspace{0.4cm}
(b)\includegraphics[angle=270,width=0.45\textwidth]{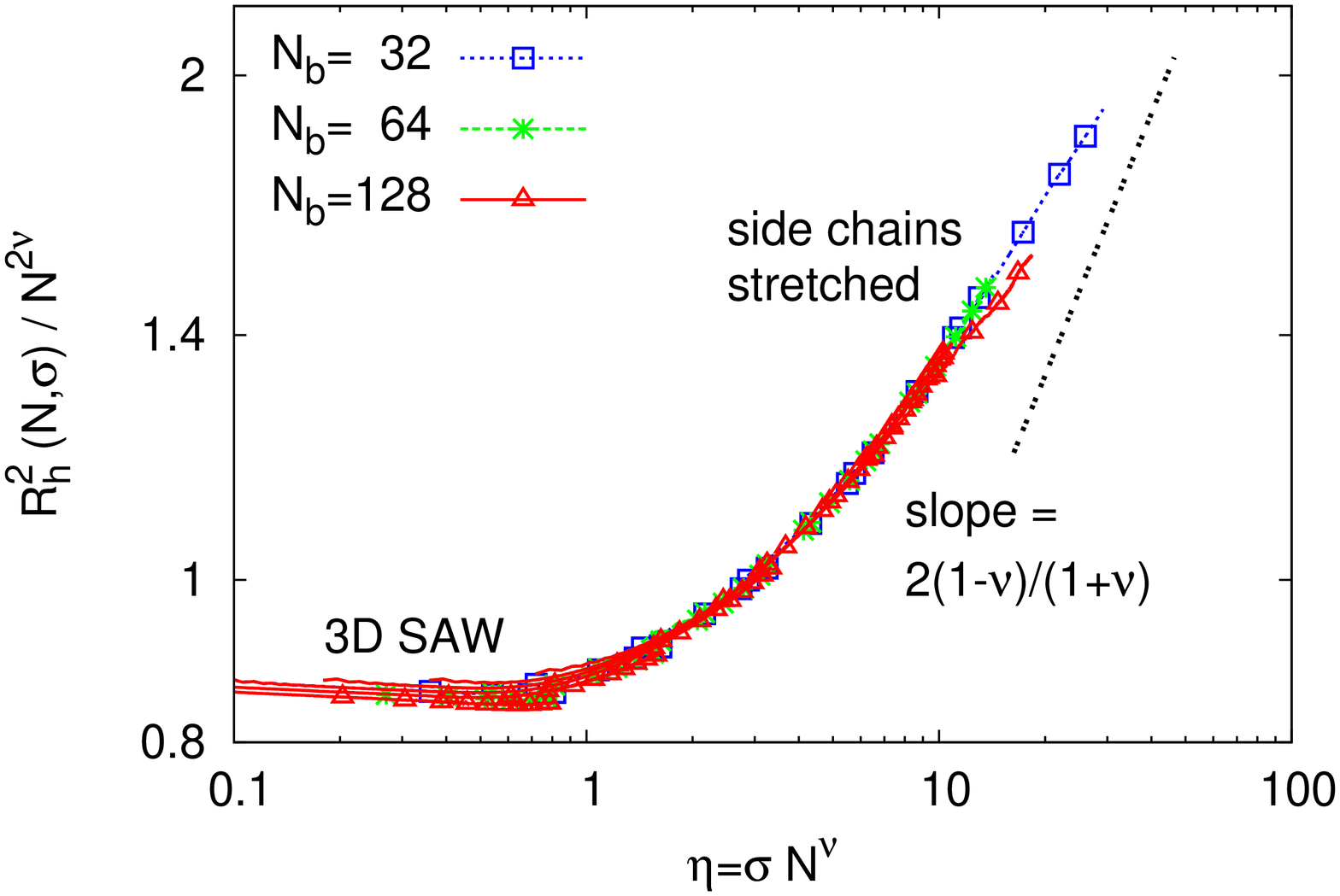}
\caption{(a) Log-log plot of rescaled mean square height
$R^2_h(N,\sigma)/N^{2\nu}$ versus $N$ (a) and $\eta=\sigma N^\nu$ (b)
with $\nu \approx 0.588$. Results are obtained for three choices of $N_b$
and several choices of the grafting density $\sigma$ as indicated.
Those unphysical data ($R_h>0.5N_b$) due to the artifact of using
periodic boundary condition are removed. The slope of the straight
line corresponds to the scaling prediction. Adapted from Ref.~\cite{Brush2}.}
\label{fig-bb-ree}
\end{center}
\end{figure}

\subsection{Bottle-brush Polymers}

   The so-called bottle-brush polymer consists of one long molecule
serving as a backbone on which many side chains are densely grafted.
As the grafting density $\sigma$ increases, the persistence length of
the backbone increases. The bottle-brush polymer has the form of
a rather stiff cylindrical-like object. If the backbone is very short
but side chains are very long, it should behave like a star polymer.
If the backbone is very long, the structure becomes more complicated.
One would expect that those side chains in the interior of the
bottle-brush are all stretched and show the same behavior,
but those at the two ends behave as a star.
In order to understand the structure of bottle-brush polymers
and check the scaling behavior of long side chains in comparison
with theoretical predictions~\cite{Brush2},
we focus here the bottle-brush polymers of a rigid backbone and
flexible side chains.

For our simulations, we use a simple coarse-grained model.
The backbone is treated as a completely rigid rod, and
side chains are described by SAWs with nearest neighbor
non-bonded attractive interactions between the same
type of monomers and repulsive interactions between
the different type of monomers.
A general formula
for the partition function for bottle-brush polymers consisting
of one or two chemically different monomers is therefore given by
\be 
    Z = \sum_{\rm config.} q^{m_{AA}+m_{BB}} q_{AB}^{m_{AB}}
\ee
where $q=\exp(-\beta\epsilon)$ (we assume that the attractive interaction
$\epsilon_{AA}=\epsilon_{BB}=\epsilon$), $q_{AB}=\exp(-\beta\epsilon_{AB})$
($\epsilon_{AB}$ is the repulsive interaction between monomer $A$ and
monomer $B$),
and $m_{AA}$, $m_{BB}$, $m_{AB}$ are the numbers of non-bonded
occupied nearest neighbor monomer pairs $AA$, $BB$ and $AB$,
respectively. For $q = 1$, all side chains behave as
SAWs. For $q<q_\Theta$ it corresponds to the good
solvent condition, where 
$q_\Theta=\exp(-\epsilon/k_BT_\Theta) \approx 1.3087$
at the $\Theta$ point~\cite{Grassberger97}. 
For $q>q_\Theta$, it corresponds to the poor solvent condition.
As $q_{AB} = 0$, it corresponds to a very strong
repulsion between $A$ and $B$, while for $q_{AB} = q$ 
the chemical incompatibility vanishes 
\{recall that~\cite{deGennes} $\chi_{AB} \propto \epsilon_{AB}-(\epsilon_{AA}
+\epsilon_{BB})/2\}$. The grafting density $\sigma$ is defined by
$\sigma=n_c/N_b$ where $n_c$ is the number of side chains and 
$N_b$ is the number of monomers in a backbone. 
Here only the results of bottle-brush polymers consisting
of one kind of monomers under a very good solvent condition
are presented in order to show the performance of the algorithm.
Other applications can be found in~\cite{Brush2,Brush1,Brush3,Brush4}.   

We extend the algorithm for simulating star polymers to
bottle-brush polymers. As shown in Fig.~\ref{fig-bottlebrush}(a), 
a bottle-brush polymer is built by adding one monomer to each
side chain at each step until all side chains have the same number
of monomers. Then we start to add the second run of monomers,
i.e, all side chains are grown simultaneously.
The bias of growing
side chains was used by giving higher probabilities in the direction
where there are more free next neighbor sites and in the outward
directions perpendicular to the backbone, where the second part of bias
decreases with the length of side chains and increases with the grafting
density. A typical configuration of bottle-brush polymers
consisting of $N_b=128$ backbone monomers, $N=2000$ side 
chain monomers, and with grafting density $\sigma=1/4$ under a 
good solvent condition is shown in
Fig.~\ref{fig-bottlebrush}(b) where the total number of monomers
is $N_{\rm tot}=128+2000 \times 32=64128$ monomers.

For checking the scaling law of side chains, we introduce the 
periodic boundary condition along the direction of the backbone
($+z$-direction) to avoid end effects associated with a finite
backbone length.
The square of the average height of a bottle-brush polymer,
$R_h^2(N,\sigma)=\langle R_{ex}^2(N,\sigma)+R_{ey}^2(N,\sigma) \rangle$ 
is estimated 
by taking the average of the mean square backbone-to-end distance 
in the radial direction for all side chains.
In Fig.~\ref{fig-bb-ree}(a) we plot $R_h^2(N,\sigma)$ 
divided by $N^{2 \nu}$ versus $N$ for $N_b=32$, $64$, and $128$,
for various values of grafting densities $\sigma$.
The value of $\nu$ is given by the best estimate for
3d SAW by PERM~\cite{Star1}.
We see that those curves of the same grafting density $\sigma$
coincide with each other.
Increasing the grafting density $\sigma$, it enhances
the stretching of side chains.
As $\sigma \rightarrow 0$, we should expect a mushroom
regime where no interaction between side chains appears.
As $\sigma$ is very high,
the scaling prediction obtained by extending the 
Daoud-Cotton~\cite{Daoud} ``blob picture"~\cite{Wang,Ligoure,Sevick,Grest95}
from star polymers to bottle-brush polymers is 
$R_h^2(N,\sigma) \propto \sigma^{2(1-\nu)/(1+\nu)} N^{4\nu/(1+\nu)}$.
Thus, we can give the cross-over scaling ansatz as follows for
$N \rightarrow \infty$,
\be 
        R^2_h(N,\sigma)= N^{2\nu} \tilde{R}^2(\eta)
\label{eq-bb}
\ee
with
\be 
    \tilde{R}^2(\eta) = \left\{ \begin{array}{ll}
1 \enspace \;, \enspace & \eta \rightarrow 0 \\
\eta^{2(1-\nu)/(1+\nu)} \enspace \; , \enspace & \eta \rightarrow \infty
\end{array}
\right .
\ee
where $\eta=\sigma N^{\nu}$. 

After removing those unphysical data
due to the artifact of using periodic boundary condition in the regime
where $R_h(N,\sigma) > N_b/2$,
we plot the same data of $R_h^2(N,\sigma)/N^{2 \nu}$
but rescaled the x-axis from $N$ to 
$\eta=\sigma N^\nu$ according to the scaling law (\ref{eq-bb}).
We see the nice data collapse in Fig.~\ref{fig-bb-ree}(b).
In this log-log plot, the straight line
gives the asymptotic behaviors of the scaling prediction 
(\ref{eq-bb}) for very large $\eta$.
As $\eta$ increases, we see a cross-over
from a 3D SAWs to a stretched side chain regime
but only rather weak stretching of
side chains is realized, which is different from
the scaling prediction.
However, this is the first time one can see the
cross-over behavior by computer simulations. 
This cross-over regime is far from reachable by experiments.

\section{PERM with Cluster Growth Method}
\label{sec-Cluster}

   It is generally believed that lattice animals, lattice trees,
and subcritical percolation are good models for studying randomly
branched polymers and they are in the same universality class.
There exist several efficient algorithms, 
e.g., Leath algorithm~\cite{Leath},
Swendsen-Wang algorithm~\cite{Swendsen}, etc. 
for studying the growth of percolation clusters
near the critical point, but they all become inefficient far below it,
because the chance for growing a large subcritical cluster by a 
straightforward algorithm decreases rapidly with $N$. Obviously we 
need some sort of cloning, and since this will probably lead also to 
fluctuating weights, one might need some pruning. 

Cloning and pruning needs first some estimate for the weight of a
cluster that is still growing. Moreover, it will turn out that 
growing clusters can have, depending on their detailed 
configurations, very different probabilities to grow further. Thus,
in addition to a weight we might to need also a ``fitness" that should
depend on the weight but is not entirely determined by it.

In the following discussion the algorithm is explained by considering the
relationship between the site percolation and site lattice animals~\cite{Anim1}.

   In any cluster growth algorithm~\cite{Leath}, a finished cluster with 
$N$ sites and $b$ boundary sites on a lattice is generated 
with probability
\be
      P_{Nb}=p^N(1-p)^b \;,
\ee
if each lattice site is occupied with the probability $p$. 
By definition of lattice animals all the clusters
of same size $N$ carry the same weight. Since the obtained
percolation cluster is biased by the probability $P_{Nb}$, 
its contribution to the animal ensemble is corrected by
a factor $1/P_{Nb}$. Taking an average over the 
percolation ensemble, the partition sum of lattice animals 
consisting of $N$ sites is given by
\be
        Z_N=\langle \frac{1}{P_{Nb}} \rangle
= p^{-N}\langle (1-p)^{-b} \rangle
\label{eq-zn-anim1}
\ee

  As shown in Fig.~\ref{fig-grow}, now we consider a cluster with $N$ sites, 
$g$ growth sites and $b$ boundary sites. At each of
the growth sites the cluster can either grow further, or it
can stop growing with the probability $1-p$. Thus, this
still growing cluster gives a weight to a percolation
cluster with $N$ sites and $(b+g)$ boundary sites
as $p^N(1-p)^{b+g}/\left[p^N (1-p)^b\right]=(1-p)^g$.
Taking an average over all clusters, we have
\be 
      Z_N= \langle \frac{(1-p)^g}{p^N(1-p)^{b+g}} \rangle
= p^{-N}\langle (1-p)^{-b} \rangle \; .
\label{reweight}
\ee
This is the same formula as given by (\ref{eq-zn-anim1}), but note that now
we have included also those clusters which are still 
{\it growing}.

\begin{figure*}
\begin{center}
\includegraphics[angle=0,width=0.60\textwidth]{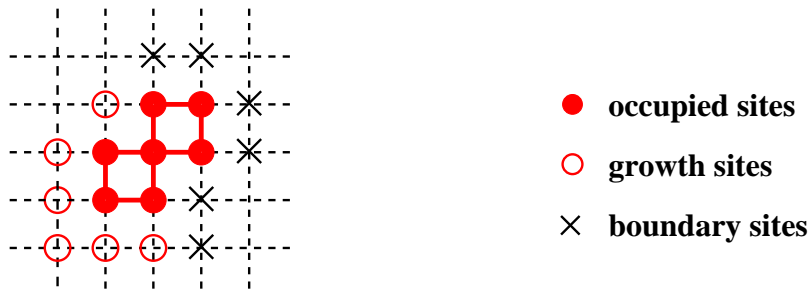}
\caption{A still growing cluster with $N=7$ sites,
$b=6$ boundary sites and $g=6$ growth sites on a square lattice.}
\label{fig-grow}
\end{center}
\end{figure*}

\begin{figure*}
\begin{center}
\includegraphics[angle=0,width=0.90\textwidth]{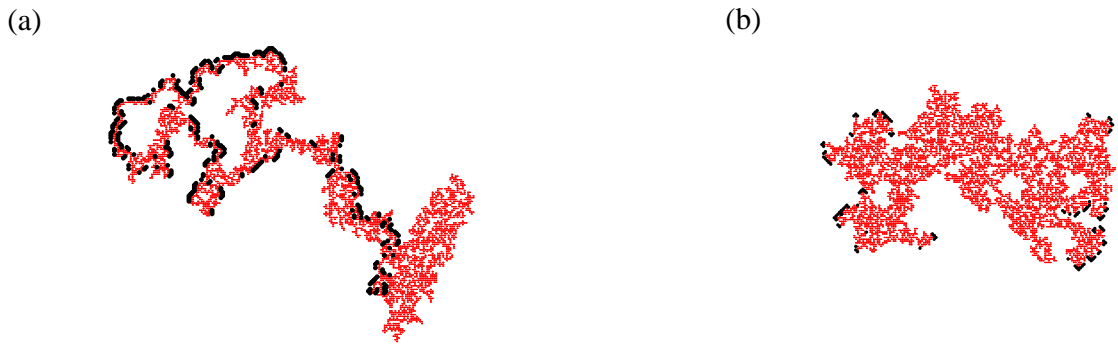}
\caption{Growing clusters generated in the (a) depth-first
and (b) breadth-first implementations. In both cases,
$p=p_c=0.5927$ and $N=4000$. Occupied sites and growth sites
are depicted by small red points and big black points,
respectively. Adapted from Ref.~\cite{Anim1}.}
\label{fig-perco}
\end{center}
\end{figure*}

  Let us first point out this new variant of PERM:

\begin{itemize}
\item The percolation cluster growth algorithm with storing
the growth sites into a queue in a first-in first-out list
(the scheme of breadth-first) is used. 

\item The population control is done by introducing a fitness
function 
\be
   f_n = W_n /(1-p)^{\alpha g} = p^{-n}(1-p)^{-b-\alpha g}      \label{fitness}
\ee
with a parameter $\alpha$ to be determined empirically, and used
\be
   f_n > c_+ \langle f_n \rangle, \qquad f_n< c_-  \langle f_n \rangle
\ee
as criteria for cloning and pruning.

\item The depth-first implementation in PERM is still used here. Namely,
at each time one deals with only a single
configuration of a cluster until a cluster has been grown either to the end of
the maximum size $N$ or has been killed in between, and
handles the copies by recursion. 

\item The optimal value of the probability $p$ is $p<p_c$, and
$p \rightarrow p_c$ as $N \rightarrow \infty$.
\end{itemize}

  This algorithm was developed more or less by trial and error, 
guided by the following considerations:

We first test the two common ways for growing the 
percolation clusters. (a) Depth-first: growth sites are written into a 
first-in last out list (a {\it stack}). (b) Breadth-first: 
growth sites are written into a first-in first-out list (a {\it queue}).
In order to avoid the mix up with the depth implementation
in PERM, we use {\it stack} and {\it queue} to distinguish
these two methods.
Two typical 2-$d$ clusters of size $N=4000$ and at the
critical point of percolation $p=p_c=0.5925$, growing 
according to these two methods are shown in 
Fig.~\ref{fig-perco}.
At first glance, one would expect that the cluster growing 
by storing growth sites in a {\it stack} might be
more efficient than that the growth sites stored in
a {\it queue}, because the number of growing sites was about 3 
times larger than that for the latter case.
But the truth is, after a few generations the descendents
generated from the former case will die. On the other hand,
the fluctuations in the number of growth sites are much bigger 
in the former case, the weights in (\ref{reweight}) will also
fluctuate much more, and we expect much worse behavior. 
This is indeed what we found numerically: 
Results obtained when using a stack for the growth sites were 
dramatically worse than results obtained with a queue.

Second, we check whether the efficiency is affected by
the chosen order of writing the neighbors of a growth site
into the list. Studying the percolation cluster in two dimensions,
one can use the preferences east-south-west-north, or 
east-west-north-south, or a different random sequence at every point. 
We found no big differences in efficiency.

Third, it would be far from optimal to do the population control as 
explained in Sect.~\ref{sec-PERM}, i.e. by using two thresholds 
$W^{\pm}$ on the current weights $W_n \equiv p^{-n}(1-p)^{-b}$. 
This would strongly favor clusters with few growth sites, since they
tend to have larger values of $b$, for the same $n$, and have thus 
large weights. But such clusters would die soon, and would thus 
contribute little to the growth of much larger clusters. Therefore a 
proper fitness function $f_n$ is needed.

Finally, we have to decide the optimal values of $p$ empirically. 
It is clear that we should not use $p>p_c$, because it is 
subcritical percolation that is in the same universality class
of lattice animal.
One might expect $p \ll p_c$ to be optimal because 
only minimal reweighting is needed for small $p$. This is 
indeed true for small $N$, but not for large $N$.
In order to reach large $N$, it is more important that clusters
grown with $p \ll p_c$ have to be cloned excessively. 
otherwise, they would die rapidly in view of their few growth sites.
In Fig.~\ref{fig-lnz-p} we present the errors of free energies $F_N=- \ln Z_N$
for various values of $p$ in $d=2$ and $d=8$.
The statistical errors always eventually decrease as 
$1/\left[{\rm CPU \, time}\right]^{1/2}$, 
hence we show there one standard deviation multiplied by 
$\left[{\rm CPU \, time}\right]^{1/2}$ 
(measured in seconds), for different values of $p$.
Thus, we can compare the accuracy between those runs on 
different computers.  
For $d=2$ (Fig.~\ref{fig-lnz-p}(a)), 
each simulation was done for $N_{\rm max}=4000$ (although we
plotted some curves only up to smaller $N$, omitting data which might not
have been converged). We see clearly that small values of $p$ are good
only for small $N$. As $N$ increases, the best results were obtained for
$p\to p_c$. The same behavior was observed also in all other dimensions,
and also for animals on the bcc and fcc lattices in 3 dimensions (data
not shown). In Fig.~\ref{fig-lnz-p}(b), we see 
the analogous results for $d=8$ and for $N_{\rm max}=8000$,
showing the errors are much smaller than those in Fig.~\ref{fig-lnz-p}(a).
Indeed, the errors decreased monotonically with $d$, being largest for $d=2$. 
Using $p$ slightly smaller than $p_c$ we can obtain easily very high 
statistics samples of animals with several thousand
sites for dimensions $\geq 2$. Another quantity which 
can help to check the reliability of our data is the tour weight
distribution (see Sect.~\ref{sec-PERM}). 
In Fig.~\ref{fig-anim-tour}, we show the two tour 
weight distributions for
two-dimensional animals with 4000 sites, for $p=0.57$ and for $p=0.47$.
We see that the simulation with $p=0.57$ is distinctly on the safe side, 
while that for $p=0.47$ is marginal. In the log-log plot, it is seen that
the tail of the distribution $P(\ln {\cal W})$ for $p=0.57$ decays faster 
than $1/{\cal W}$, thus the product ${\cal W}P(\ln {\cal W})$ has its 
maximum where the distribution is well sampled.

Error bars quoted in the following on raw data (partition sums,
gyration radii, and average numbers of perimeter sites or bonds) are
straightforwardly obtained single standard deviations. Their estimate
is easy since clusters generated in different tours are independent,
and therefore errors can be obtained from the fluctuations of the
contributions of entire tours (notice that clusters within one tour
are {\it not} independent, and estimating errors from their individual
values would be wrong).

\begin{figure}
\begin{center}
(a)\includegraphics[angle=270,width=0.45\textwidth]{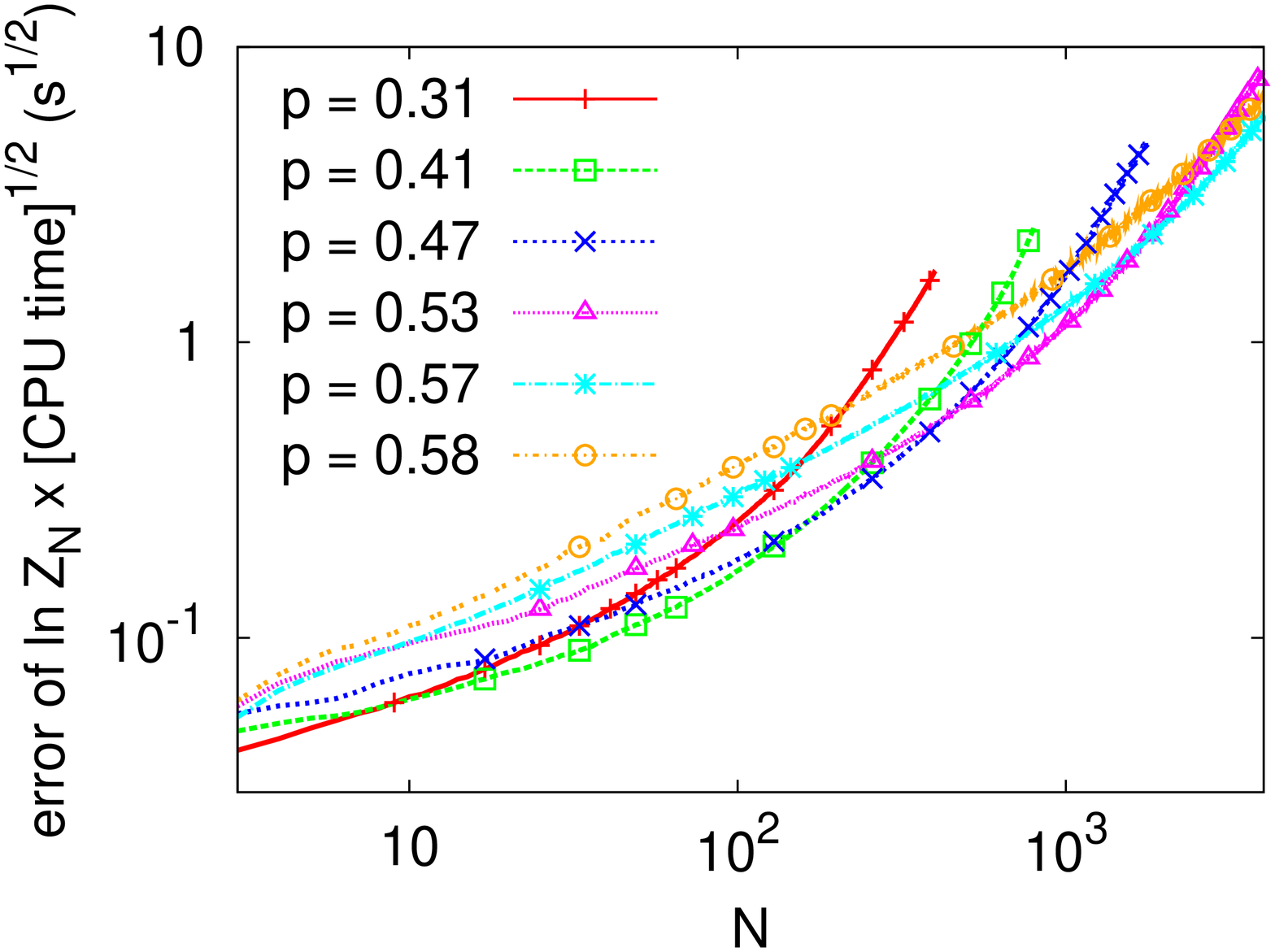}\hspace{0.4cm}
(b)\includegraphics[angle=270,width=0.45\textwidth]{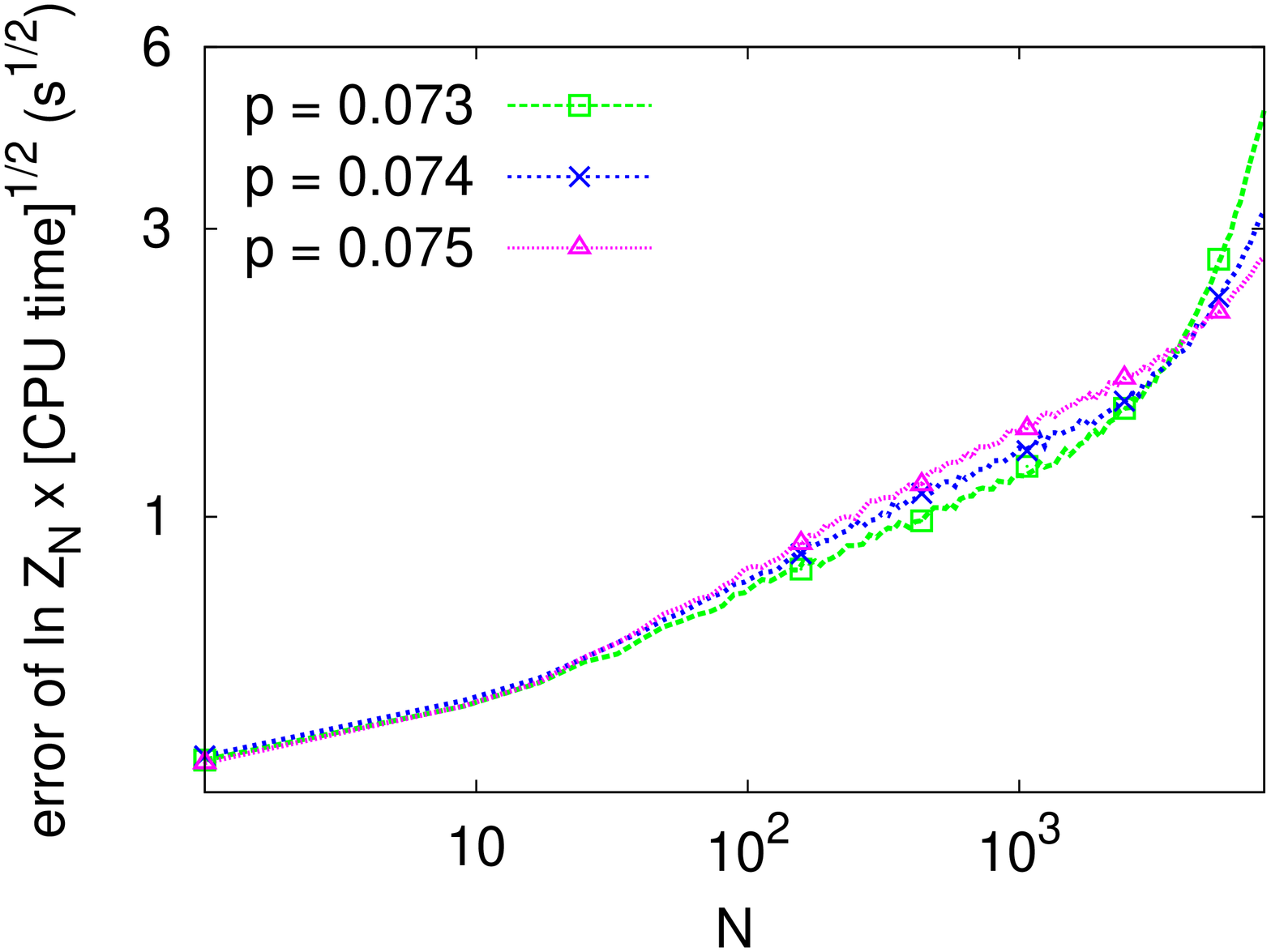}
\caption{Statistical errors of $\ln Z_N$ for lattice animals
in $d=2$ (a) and $d=8$ (b) for various values of $p$.
To make the different runs comparable, errors are multiplied by the
square root of the CPU time measured in seconds.
The cluster size $N$ is up to $4000$ in (a) and up to $8000$ in (b).
The percolation thresholds are $p_c=0.5927$ in $d=2$, and $p_c=0.0752$ 
in $d=8$. Adapted from Ref.~\cite{Anim1}.}
\label{fig-lnz-p}
\end{center}
\end{figure}

\begin{figure}
\begin{center}
\includegraphics[angle=270,width=0.45\textwidth]{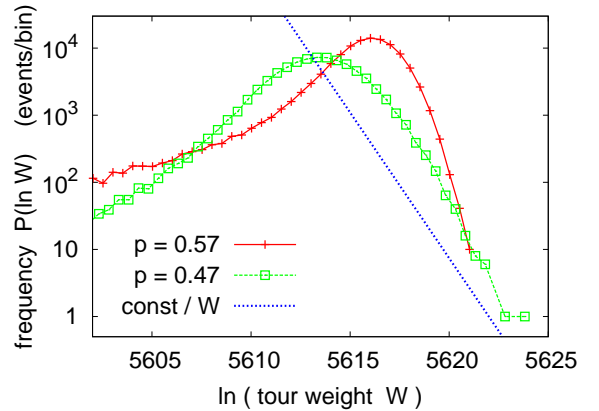}
\caption{Log-log plot of distributions of tour weights
$P(\ln W)$ of $2d$ animals with $N=4000$, for
$p=0.57$ and $p=0.47$, together with a straight line representing
the function $y=const/W$. Adapted from Ref.~\cite{Anim1}.}
\label{fig-anim-tour}
\end{center}
\end{figure}

\begin{figure*}[htb]
\begin{center}
\includegraphics[angle=0,width=0.80\textwidth]{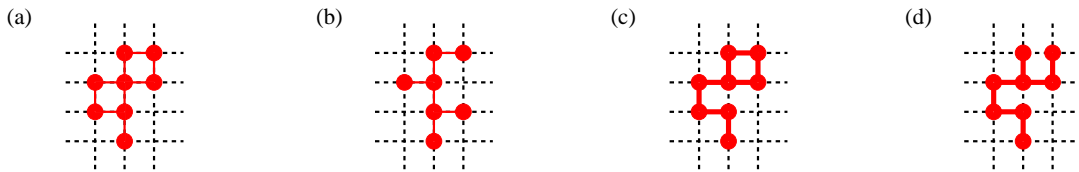}
\caption{(a) A site animal with $8$ sites. (b) A site tree (``strongly embeddable tree").
(c) A bond animal which is not a tree. (d) A bond tree (``weakly embeddable tree").}
\label{fig-anim-tree}
\end{center}
\end{figure*}

\begin{figure*}[htb]
\begin{center}
\includegraphics[angle=0,width=0.80\textwidth]{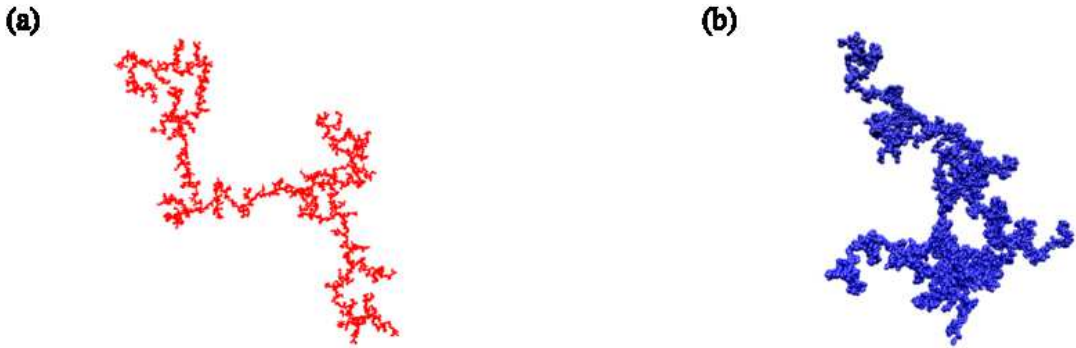}
\caption{Typical site lattice animals with $N=12000$ on the square
lattice in $d=2$ (a), with $N=16000$ on the bcc lattice in $d=3$ (b).}
\label{fig-anim2d3d}
\end{center}
\end{figure*}

In addition to site animals, this algorithm can also be
applied to bond animals and lattice trees for studying randomly 
branched polymers.
A bond animal is a cluster where bonds can be established between
neighboring sites (just as in SAWs), and connectivity is defined via
these bonds: if there is no path between any two sites consisting entirely of
established bonds, these sites are considered
as not connected, even if they are nearest neighbors.
Different configurations of bonds are considered as different clusters,
and clusters with the same number of bonds (irrespective of
their number of sites) have the same weight~\cite{Janse}.
Weakly embeddable trees are bond animals
with tree topology, i.e. the set of weakly embeddable trees is a subset of
bond animals, each with the same statistical weight. Strongly 
embeddable trees are,
in contrast, the subset of site animals with tree-like structure.
All these definitions are illustrated in Fig.~\ref{fig-anim-tree}.

\subsection{Non-interacting Lattice Animals in the Bulk}

    The basic problem of lattice animals ({\it site animals})
is how to count the number of different animals of $N$ sites precisely,
i.e. the estimate of the
corresponding partition sum. Two animals are considered as identical
if they differ just by a translation, but they considered as different
if a rotation or reflection is needed to make them coincide.
Two typical site animals consisting of $N=12000$ sites on the square
lattice in $d=2$ and with $N=16000$ sites on the body centered cubic
(bcc) lattice in $d=3$ are shown in Fig.~\ref{fig-anim2d3d}.

In the thermodynamic limit as $N \rightarrow \infty$, the number
of animals (i.e. the microcanonical partition sum) should scale
as~\cite{Lubensky}
\begin{equation}
     Z_N \sim \mu^{N} N^{-\theta}(1+b_zN^{-\Delta}+\cdots) \;,
\label{eq-zn-anim}
\end{equation}
and the gyration radius as
\begin{equation}
    R_N \sim N^{\nu}(a+b_R N^{-\Delta}+\cdots)
\label{eq-rn-anim}
\end{equation}
Here $\mu$ is the growth constant (or {\it inverse critical fugacity}),
and is not universal, while the Flory exponent $\nu$, the entropic
exponent $\theta$, and the correction exponent $\Delta$~\cite{Adler}
should be universal. $b_z$ and $b_R$ are non-universal amplitudes,
and the dots stand for higher order terms in $1/N$.

Results of the partition sum $Z_N$ and the mean square end-to-end
distance $R^2_N$ for site lattice animals in $d=2$ are shown
in Fig.~\ref{fig-anim2d}. By taking the predicted value of $\theta=1$
and plotting $\ln Z_N-aN+\ln N$ against $N$, we should expect a curve
which becomes horizontal for large $N$ by adjusting values of $a=\ln \mu$
suitably. This is indeed seen for the central curve with error bar
in the inset of Fig.~\ref{fig-anim2d}(a),
but a precise estimate of $\mu$ is difficult because of corrections
to scaling. Considering the first correction term in
(\ref{eq-zn-anim}) and (\ref{eq-rn-anim}), the correction exponent $\Delta$,
and the estimate of the growth constant $\mu$ and their error bars
are all determined by the best straight line as $N^{-\Delta} \rightarrow 0$ in
Fig.~\ref{fig-anim2d}. Our estimate of $a=\ln \mu =1.4018155(30)$ with
$\Delta=0.9(1)$ is in perfect agreement with the exact enumeration
result~\cite{Jensen}. The Flory exponent $\nu$ is determined by
the same way and our estimate $\nu=0.6412(5)$ is also in good agreement
with the previous estimate by Monte Carlo simulations~\cite{You}.

\begin{figure}
\begin{center}
(a)\includegraphics[angle=270,width=0.45\textwidth]{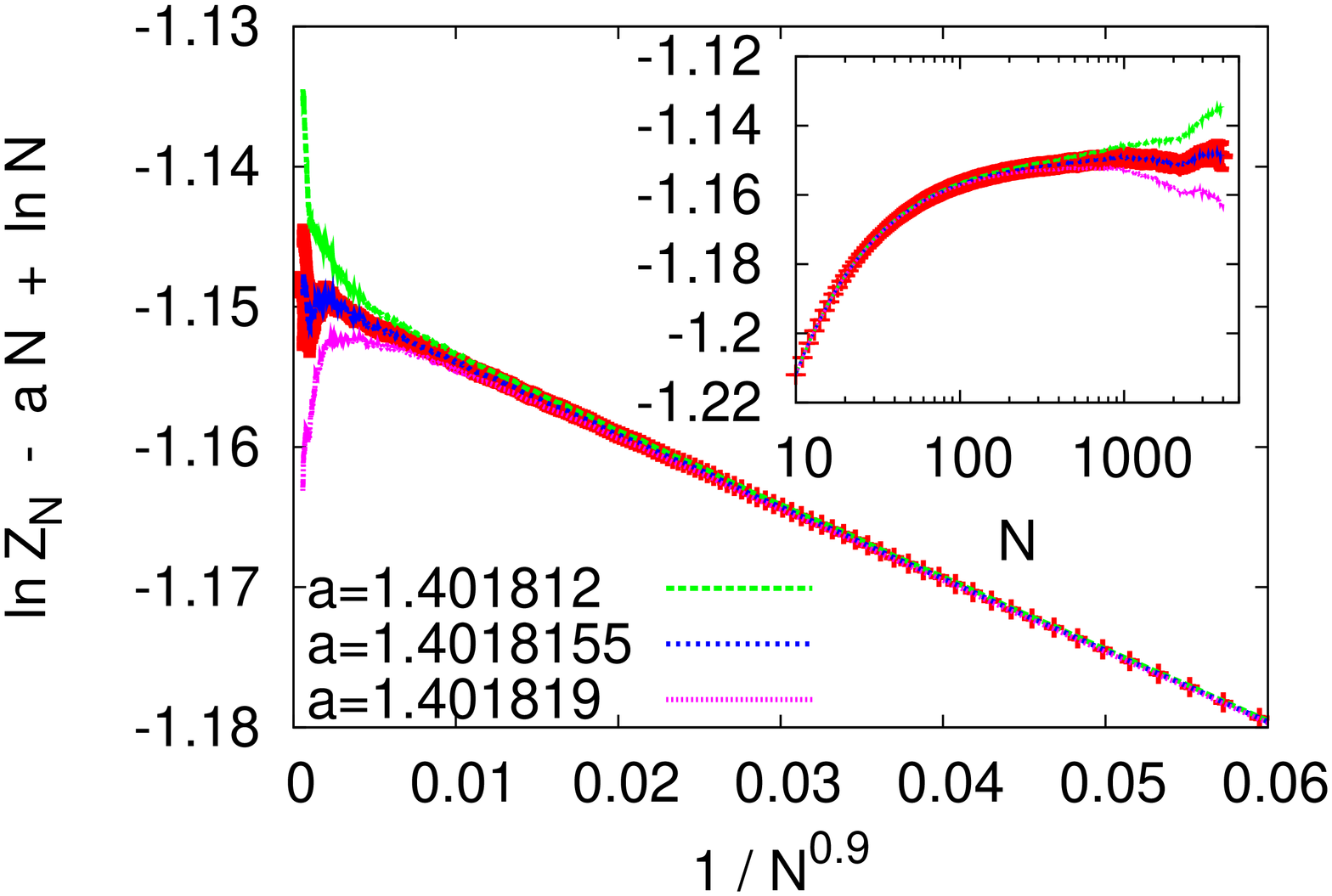}\hspace{0.4cm}
(b)\includegraphics[angle=270,width=0.45\textwidth]{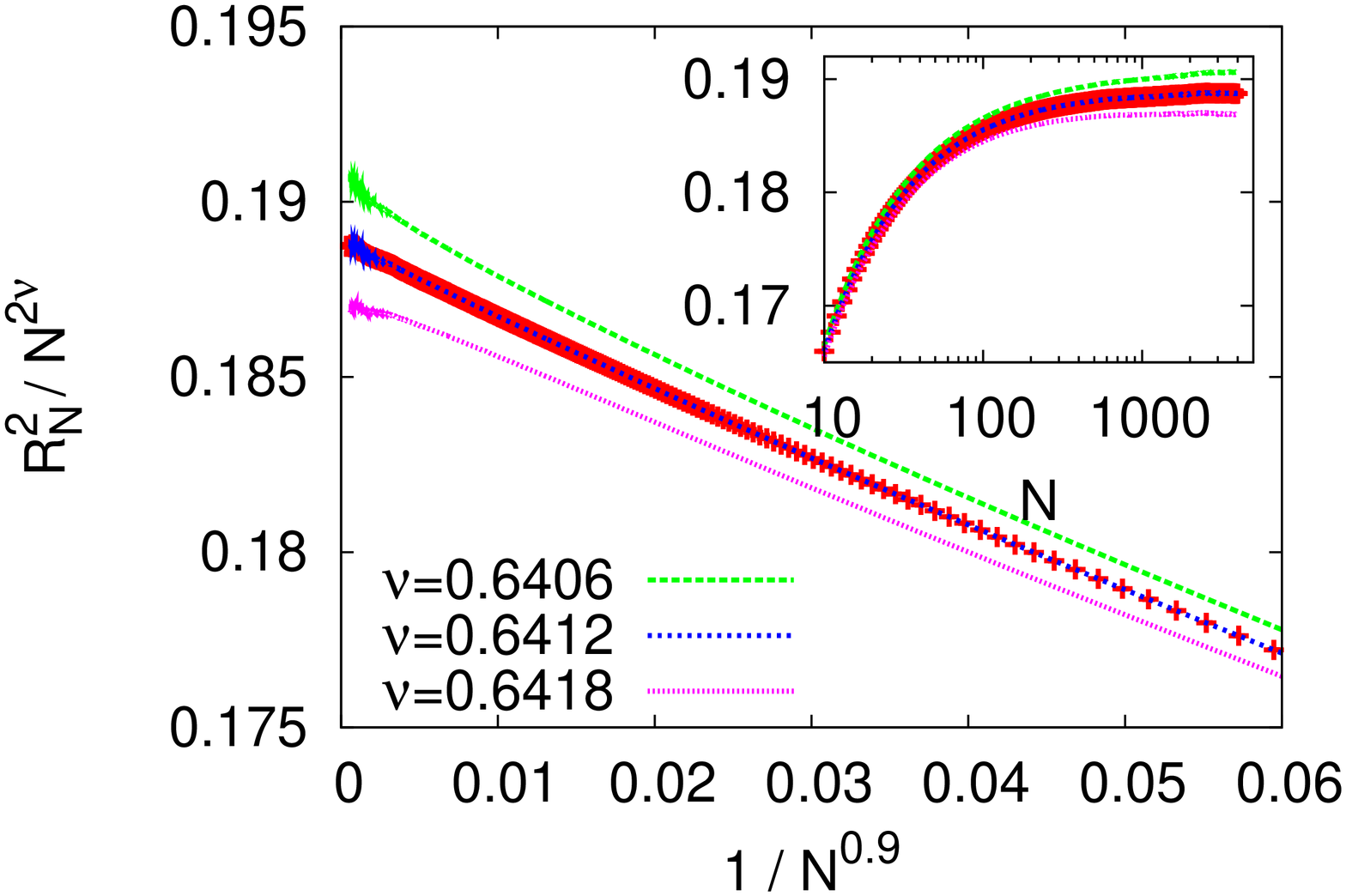}
\caption{(a) Results of $\ln Z_N-aN+\theta \ln N$ plotted against $N^{-\Delta}$,
and against $N$ (in the inset), and (b) results of $R^2_N/N^{2\nu}$ plotted 
against $N^{-\Delta}$, and against $N$ (in the inset). 
The best estimates of $a=\ln \mu=1.4018155(30)$, $\nu=0.6412(5)$ and $\Delta=0.9(1)$ 
are given by the best straight lines. All data are for site lattice animals in $d=2$. Adapted from Ref.~\cite{Anim1}.}
\label{fig-anim2d}
\end{center}
\end{figure}

It is trivial to generalize the algorithm PERM with cluster growth
method to lattice animals in higher dimensions $2 < d \le 9$.
Using the similar method of
data analyses as shown in Fig.~\ref{fig-anim2d} for those results
obtained in $d=3$ to $d=7$ ($d_c=8$ is the upper critical
dimension of lattice animals, where large corrections have to be
taken into account). The relationship between the entropic exponent $\theta$
and the Flory exponent $\nu$ for the animal problem in $d$ dimensions is
predicted by using supersymmetry~\cite{Parisi},
\be
    \theta=(d-2)\nu+1
\label{eq-Parisi}
\ee
By plotting our data of the exponent $\nu$ and $(\theta-1)/(d-2)$
against $d$ in Fig.~\ref{fig-Parisi},
we see that these two curves coincide with each other. It shows that
the Parisi-Sourlas prediction (\ref{eq-Parisi}) is verified.

\begin{figure}
\begin{center}
\includegraphics[angle=270,width=0.45\textwidth]{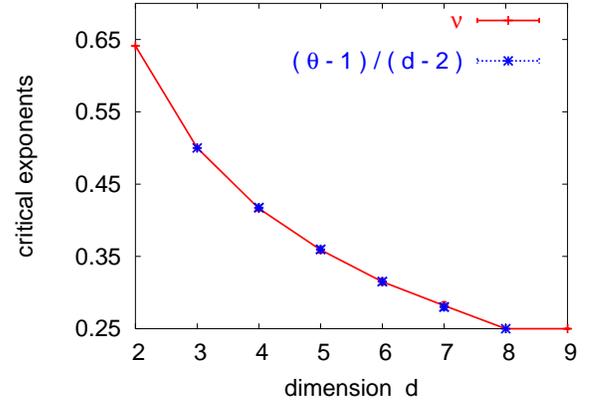}
\caption{The critical exponents $\nu$ and $(\theta-1)/(d-2)$ 
against $d$. Adapted from Ref.~\cite{Anim1}.}
\label{fig-Parisi}
\end{center}
\end{figure}

\subsection{Lattice Animals Grafted to Surfaces}

For a-thermal walls (which represent only a geometric barrier, without
any other interactions) the leading behavior for $N\to\infty$ does not involve 
any new critical exponent \cite{Anim1}. This is no longer true, however,
if the wall is attractive. In that case we expect a phase transition at 
a critical attractive energy beyond which the animal gets adsorbed to 
the surface, similar to the adsorption transition observed also for linear
polymers \cite{Graf-G05}. 

As in that problem, at the transition point there are new critical exponents.
More precisely, the Flory exponent $\nu$ is the same as for non-grafted animals,
but the entropic exponent $\theta$ is changed \cite{Anim1}. Since this 
exponent could not be measured by any previous simulation algorithm and since 
there exits no field-theoretic predictions for it, there exist no literature
values to compare to our measurements. This is different for a second new exponent
specific for the transition point, the cross-over exponent $\phi$. If $q$ is 
the Boltzmann factor for the monomer-wall interaction and $q_c$ is its critical
value, then the scaling ansatz for the partition sum of a grafted animal near 
the adsorption transition is 
\be
   Z^{(1)}_N(q) \sim \mu(q)^N N^{-\theta_s} \Psi[(q-q_c)N^\phi].
\ee
The most interesting prediction for $\phi$ was that is {\it superuniversal}, i.e.
its value is independent of the dimension and $\phi=1/2$ for all dimensions
\cite{Lyssy}. While this was verified by the simulations for $d=3, 4$ and 5, 
it was slightly violated (by 5 standard deviations) in $d=2$ \cite{Anim1}. 
Obviously further investigations would be needed to settle this problem.

\subsection{Conformal Invariance and Animals Grafted to Wedges}

The critical exponents for animals in $d=2$ dimensions can be calculated exactly,
as for many other critical phenomena in $d=2$ dimensions. But while this is due 
to conformal invariance in these other cases, 2-d animals are not conformally 
invariant \cite{Miller}. 

\begin{figure}
\begin{center}
\includegraphics[angle=270,scale=0.28]{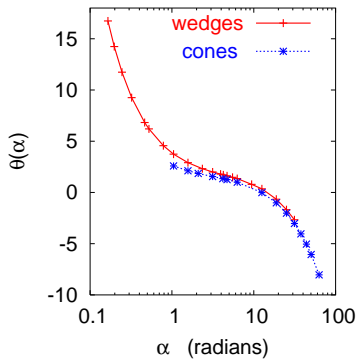}
\caption{Entropic critical exponents $\theta(\alpha)$ for 2-d lattice animals
   grafted to the tips of wedges resp. cones with angles $\alpha$.
Adapted from Ref.~\cite{Anim3}.}
\label{fig-wedges}
\end{center}
\end{figure}

For conformally invariant problems of cluster growth,
the entropic critical exponents of clusters grafted to the tips of wedges and 
cones (wedges with identified edges) can be calculated exactly for any wedge
angle, by mapping the wedge onto the half plane. Due to lack of conformal
invariance, this is no longer true for 2-d lattice animals. In \cite{Anim3},
the exponents $\theta(\alpha)$ were measured carefully not only for wedges and cones
with angles up to $\alpha=2\pi$. By grafting them to branch points of Riemann
sheets, angles up to $10\pi$ were studied. Results are shown in Fig. \ref{fig-wedges}.
The simulations were made with the hope that someone might produce a fit to these 
data that could suggest an alternative to conformal invariance. So far this hope
has not materialized, in contrast to what happened 111 years ago to some obscure
black body radiation data \cite{Lummer}.

\subsection{Collapsing Lattice Animals and Lattice Trees in $d=2$}

A coil-globule transition similar to that for linear polymers is also expected to 
occur for randomly branched polymers as the solvent quality becomes worse, but the 
situation is much more complicated. To describe the possible collapse transitions
for self-interacting lattice animals, we need two different types of interactions
between nearest-neighbor monomer-monomer pairs: (covalent) bonds that are needed 
for the connectedness of the cluster but that can also form loops when present in
excess, and weak interactions between non-bonded pairs (``contacts"). Associated 
to these are two different control parameters \cite{Flesia}.
The partition sum is therefore written as follows~\cite{Anim2}
\be
      Z_N(y,\tau)=\sum_{b,k} C_{Nbk} y^{b-N+1} \tau^k
\ee
where $C_{Nbk}$ is just the number of configurations (up to translations
and rotations) of connected clusters with $N$ sites, $b$ bonds, and $k$ contacts.
$y$ and $\tau$ are fugacities for monomer-monomer bonds and for non-bonded
monomer-monomer contacts, respectively. As for unbranched polymers,
there is no need to introduce a separate monomer-solvent interaction,
since the number $s$ of monomer-solvent contacts is not independent,
but is given by
\be
    {\cal N}N = 2b+2k+s \;,
\ee
where ${\cal N}$ is the lattice coordination number (${\cal N}=2d$ on a
simple hypercubic lattice in $d=2$ dimensions). A schematic drawing
of an interacting lattice animal is shown in Fig.~\ref{fig-cmodel}.
This model includes the following special cases:
\begin{itemize}
\item Unweighted animals: $y=\tau=1$
\item Bond percolation: $y=p/(1-p)^2$ and $\tau=1/(1-p)$ where
$0 \le p \le 1$. The critical percolation point is at $y=2$ and $\tau=2$
as $p=p_c=1/2$.
\item Collapsing trees: $y=0$ where $b=N-1$.
\item `Strongly embeddable' animals with $\tau=0$, which have no contacts ($k=0$)
This model was first studied by Derrida and Herrmann by transfer
matrix methods \cite{Derrida}.
\end{itemize}

\begin{figure*}
\begin{center}
\includegraphics[angle=0,width=0.75\textwidth]{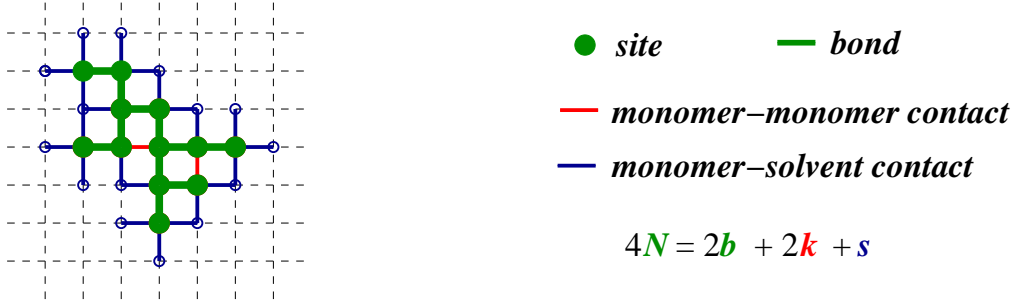}
\caption{A schematic drawing of an interacting lattice animals
which contain a cluster with $b=11$ bonds and $N=12$ sites, and
$k=2$ non-bonded monomer-monomer contacts, and $s=22$.
It leads to $4N=2b+2k+s$.}
\label{fig-cmodel}
\end{center}
\end{figure*}

\begin{figure*}
\begin{center}
\includegraphics[angle=270,width=0.60\textwidth]{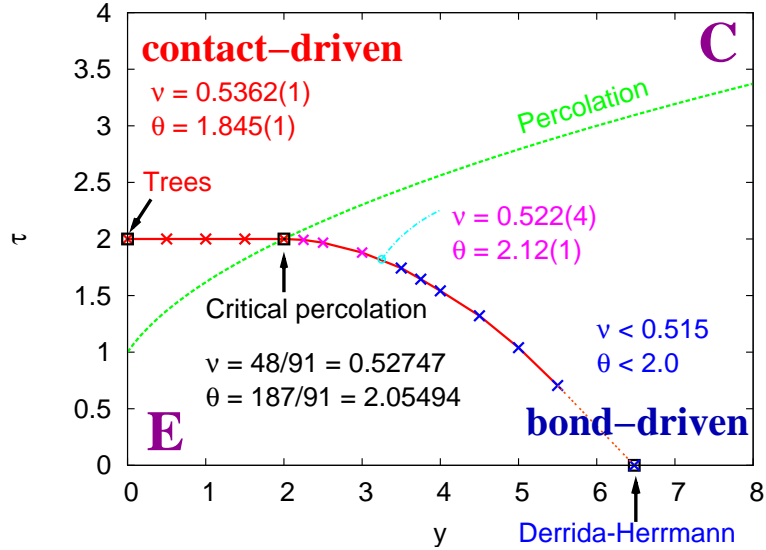}
\caption{Phase diagram for interacting animals in $d=2$.
The full curve separates an extended phase (below) from a
collapsed phase (above). At $y=0$ the clusters are
trees (minimal number of bonds), while at $\tau=0$
they have no contacts but only bonds. The dashed line
corresponds to bond percolation, with the critical point
being at $y=\tau=2$. The short dashed-dotted line is
a rough estimate for a possible transition between a contact-rich
and a bond-rich collapsed phase.
The critical exponents $\nu$ and $\theta$ are also shown
for different universality classes. Adapted from Ref.~\cite{Anim2}.}
\label{fig-canim2d}
\end{center}
\end{figure*}

   The transition points are determined by the scaling laws of the
partition sum:
\be
      Z_N(y,\tau=\tau_c) \sim \mu(y)^N N^{-\theta}
\ee
and the gyration radius
\be
      R_N \sim N^\nu
\ee
where $\mu(y)$ should depend continuously on $y$, but $\theta$ should
take discrete values depending on the respective universality.
$\nu$ is the Flory exponent. A phase diagram for interacting animals in $d=2$
is shown in Fig.~\ref{fig-canim2d}. Lattice animals are in the extended phase
below the full line, but they are in the collapsed phase above the full line.
The bond percolation is described by the dashed curve.
The percolation critical point at $y=2$ and $\tau=2$ divides the transition
line into two different universality classes.
On the left-hand side, the collapse transitions are dominated by
non-bonded contacts. In this region PERM simulations are very easy and yield 
very precise values for the transition curve (which seems to be exactly 
horizontal) and for the critical exponents. These results have been fully
confirmed by field theoretic methods \cite{Janssen-anim}. For the 
Derrida-Herrmann model at the far right end
of the transition curve, PERM simulations are least efficient, and 
they could not improve on the results of \cite{Derrida}.
It is not entirely clear whether there exists a further (multi-)critical point
between this end point and the percolation point. Such a point, together with
an additional phase separation line emanating from it, was suggested by 
earlier exact enumeration studies (cited in \cite{Anim2}). PERM simulations 
also weakly suggested such an addition phase separation line, indicated by
the short dashed-dotted line in Fig. \ref{fig-canim2d}, but these simulations 
were not easy and interpreting their results was not unambiguous. Indeed, 
a completely different scenario for the behavior along the transition line 
between the  percolation and Derrida-Herrmann points is suggested 
in \cite{Janssen-anim}.

\section{Protein Folding}

In this section we shall only describe applications of a variant of 
PERM~\cite{NPERM1,NPERM2}
to simple lattice models, where it seems one of the most efficient
algorithms for finding low energy states. PERM was also applied to 
continuum models~\cite{NPERM-AB} and was there more efficient than previous
Monte Carlo algorithms~\cite{Stillinger}, but has been rendered later obsolete 
in this application~\cite{Bachmann-Arkin}.

\subsection{New Version of PERM (nPERM)}
\label{sec-nPERM}
  The main improvement of nPERM is that we no longer make 
{\it identical clones} as implemented in old PERM,
in order to avoid the loss of diversity which limited the 
success of old PERM.

When we have a configuration of polymer chains with $n-1$ monomers, 
we first estimate a {\it predicted weight} $W_n^{\rm pred}$
for the next step (the $n^{\rm th}$ step), and we count
the number $k_{\rm free}$ of free sites where the $n^{\rm th}$
monomer can be placed. If  $W_n^{\rm pred} > W_n^+$ and $k_{\rm free}>1$,
we make $k$ ($2 \leq k \leq k_{\rm free}$) clones
with the request that 
$k$ different sites are chosen for the $n^{\rm th}$ step.
Therefore, $k$ configurations with $n$ monomers are {\it forced} 
to be different. If $W_n^{\rm pred} < W_n^-$, a random number $r$
is chosen uniformly in $[0,1]$. If $r<1/2$, the chain is discarded,
otherwise it is kept and its weight is doubled. We tried several
strategies for selecting $k$ which all gave similar results.
Typically, we used 
$k={\rm min}\left\{k_{\rm free}, \left[W_n^{\rm pred}/W_n^+\right]\right\}$.

It is still important to keep the right weight of each configuration
with $n$ monomers.
When selecting a $k$-tuple $A=\left\{\alpha_1, \ldots ,\alpha_k\right\}$
of mutually different continuations $\alpha_j$ with probability $p_A$,
the corresponding weights $W_{n,\alpha1}$, $\ldots$, $W_{n,\alpha_k}$
are 
\be
   W_{n,\alpha_j} = \frac{ W_{n-1} q_{\alpha_j} k_{\rm free}}
                     { k {k_{\rm free}\choose k} p_A}\;,
 \enspace \enspace j=1,2,\ldots,k \; .
\label{nPERMW}
\ee
Here, the {\it importance}
\be
   q_{\alpha_j}=\exp(-\beta E_{n,\alpha_j})
\ee
of choice $\alpha_j$ is the Boltzmann-Gibbs factor associated with
the energy $E_{n,\alpha_j}$ of the $n^{\rm th}$ placed monomer
in the potential created by all previous monomers.
The other terms arise from correcting bias and normalization.

Two strategies for the choice of $k$ continuations among
$k_{\rm free}$ are described as follows:

\begin{enumerate}
\item[(i)] New PERM with simple sampling (nPERMss):\\
$k$ different free sites are chosen randomly and
uniformly. The predicted weight is given by,
\be
      W_n^{\rm pred} = W_{n-1} k_{\rm free} \;,
\ee
and the corresponding weight for each continuation $\alpha_j$ is
\be
      W_{n,\alpha_j} = W_{n-1} q_{\alpha_j} k_{\rm free}/k \; 
\ee
since there are ${k_{\rm free}\choose k}$ different ways to
select a $k$-tuple with equal probability, the probability $p_A$ 
is therefore
\be
       p_A={k_{\rm free}\choose k}^{-1} \; .
\ee
Here the tuples related by permutations are considered as identical.
\item[(ii)] New PERM with importance sampling (nPERMis):\\
$k$ different free sites are chosen according to the 
modified Boltzmann weight $\tilde{q}_{\alpha_j}$ defined by
\be
 \tilde{q}_{\alpha_j}=(k_{\rm free}^{(\alpha_j)}+1/2)
\exp(-\beta E_{n,\alpha_j})
\ee
where $k_{\rm free}^{(\alpha_j)}$ is the number of free neighbors 
when the $n^{\rm th}$ monomer is placed at $\alpha_j$,
and $E_{n,\alpha_j}$ is its energy gain. The idea of replacing
$q_{\alpha_j}$ by $\tilde{q}_{\alpha_j}$ is that we anticipate
continuations with fewer free neighbors which will contribute
less on the long run than continuations with more free neighbors.
This is similar to ``{\it Markovian anticipation}" within
the framework of old PERM, described in section Sect.~\ref{sec-Markovian},
where the bias for placing a monomer at the next step different 
from the short-sighted optimal importance sampling was found
to be preferable. The predicted weight is now
\be 
     W_n^{\rm pred} = W_{n-1} \sum_{j=1}^{k_{\rm free}} 
\tilde{q}_{\alpha_j} \;.
\ee
Using the requirement that the variance of the weights $W_n$ is 
minimal, the proper choice of the probability $p_A$ to select a 
tuple $A=\left\{\alpha_i, \ldots ,\alpha_k\right\}$ is found to be
\be
     p_A=\frac{\sum_{\alpha_j \in A}\tilde{q}_{\alpha_j}}
      {\sum_{A'} \sum_{\alpha'_j \in A'} \tilde{q}_{\alpha'_j} }\;,
 \enspace \enspace j =1,2,\dots,k_{\rm free} \;.
\ee
If $q_{\alpha_j}$ had not been replaced by $\tilde{q}_{\alpha_j}$,
the variance of $W_n$ for fixed $W_{n-1}$ would be zero.
For $k=1$, it corresponds to the standard importance sampling, i.e.
$p_A=p_{\alpha_j}=\tilde{q}_{\alpha_j}/\sum_{i=1}^{k_{\rm free}}
\tilde{q}_{\alpha_i}$. The weight at the $n^{\rm th}$ step is thus
$W_{n,{\alpha_j}}=W_{n-1} q_{\alpha_j}/p_{\alpha_j}$.
For $k>1$, $W_{n,{\alpha_j}}$ is given by (\ref{nPERMW}).

\end{enumerate}

     A noteworthy feature of both nPERMss and nPERMis is 
that they cross over to complete enumeration when $W_n^+$
and $W_n^-$ tend to zero. In this limit, all possible branches
are followed and none is pruned as long as its weight is not strictly
zero. In contrast to this, with the use of the original PERM as explained in 
Sect.~\ref{sec-PERM}, exponentially many copies of the same configuration
would be made. It suggests that one can be more lenient in
choosing $W_n^+$ and $W_n^-$ when applying nPERM.


\subsection{HP Model}

 For testing the efficiency of the new PERM, we applied it to
the HP model \cite{Dill} since this model is well simulated for bench-marking.
In this model, a protein is simplified by replacing amino acids
by only two types of monomers, $H$ (hydrophobic)
and $P$ (polar) monomers. Therefore a protein (a polymer) of length $n$ 
is modeled as a self-avoiding chain of $n$ steps on a regular 
(square or simple cubic) 
lattice with repulsive or attractive interactions between 
neighboring non-bonded monomers
such that $\epsilon_{HH}=-1$, $\epsilon_{HP}=\epsilon_{PP}=0$.
The partition sum is 
\be
        Z_n=\sum_{walks} q^m
\ee
where $q=\exp(-\beta \epsilon_{HH})$ and $m$ is the total number
of non-bonded $H-H$ pairs.

In our simulations, we chose the two thresholds $W_n-=0$ and
$W_n^+ \leq \infty$, i.e. we neither pruned nor branched,
for the first configuration hitting length $n$.
For the following configurations we used 
$W_n^+=C\hat{Z}_n/\hat{Z}_0 (c_n/c_0)^2$
and $W_n^-=0.2\,W_n^+$. Here, $c_n$ is the total number of 
configurations of length $n$ already created during the run, 
$\hat{Z}_n$ is the partition sum estimated from these 
configurations, and $C$ is some positive number $\le 1$.
The idea of incorporating the term $(c_n/c_0)^2$ is that
we can reduce the upper threshold $W_n^+$ in order to make
more cloning in possible branches as a lower energy state
is hit but only few configurations of length $n$ have been obtained.  
The following results were all obtained with $C=1$, though substantial
speed-ups (up to a factor 2) could be
obtained by choosing $C$ much smaller, typically as small as $10^{-15}$ to
$10^{-24}$. The latter is easily understandable: with such small $C$, the
algorithm performs essentially exact enumeration for short chains, giving
thus maximal diversity, and becomes stochastic only later when following
all possible configurations would become unfeasible. 

  For presenting the efficiency of nPERMss and nPERMis, we 
applied them to find the ground state of the HP model with
blind search. Special comparison is made with the
{\it core-directed growth method} (CG) of 
Beutler and Dill~\cite{bd96}. This is the only method we
found to be still competitive with nPERM but it works only
for the HP model and relies heavily on heuristics.
Two examples are shown here.

{\bf (a)} Ten sequences of 48-mers in $d=3$ from Ref.~\cite{yue95}
are tested. In Table~\ref{table1}, we list the required CPU time
(measured in minutes) per independent ground state hit on  
a 167 MHz Sun ULTRA I workstation. As with the original 
PERM~\cite{Grassberger97}, we could also reach lowest energy states by using
nPERM, but the required CPU time is within one order of magnitude 
shorter than that needed for PERM. For all ten sequences we
use the same temperature, $\exp(1/T)=18$, although we could
have optimized CPU times by using different temperatures
for each chain.  
Results obtained in Ref.~\cite{NPERM1,NPERM2} are carried out
on a SPARC 1 machine which is slower by a factor $\approx 10$ than
the Sun workstation. Therefore, in Table~\ref{table1}
we multiplied their results by 10 for comparison. 
We see that nPERM gave comparable speeds as CG.
But, one has to note that the lowest energy of the sequence No.9 was 
not hit by CG~\cite{bd96}. 

\begin{table}
\caption{Performances for the 3-d binary (HP)- sequences of 48-mers from
\protect~\cite{yue95}, presented by the CPU time (minutes) per independent
ground state hit. The ground state energy is denoted by $E_{\min}$.
Results obtained by using PERM~\cite{PS321998}, nPERMss  
and nPERMis~\cite{NPERM1,NPERM2} are carried out 
on a 167 MHz Sun Ultra I workstation. Results quoted from 
Ref.~\cite{bd96} obtained by CG are multiplied by 10
for the comparison.}
\label{table1}
\begin{center}
\begin{tabular}{ccrrrr}
\hline
 sequence nr. & $-E_{\rm min}$ & PERM & nPERMss & nPERMis  & CG\\ \hline
    1     &    32    &    6.9  &   0.66 &   0.63  &  0.94 \\
    2     &    34    &   40.5  &   4.79 &   3.89  &  3.50 \\
    3     &    34    &  100.2  &   3.94 &   1.99  &  6.20 \\
    4     &    33    &  284.0  &  19.51 &  13.45  &  2.90 \\
    5     &    32    &   74.7  &   6.88 &   5.08  &  1.20 \\
    6     &    32    &   59.2  &   9.48 &   6.60  & 46.00 \\
    7     &    32    &  144.7  &   7.65 &   5.37  &  6.40 \\
    8     &    31    &   26.6  &   2.92 &   2.17  &  3.80 \\
    9     &    34    & 1420.0  & 378.64 &  41.41  &  -    \\
   10$\;\;$ &  33    &   18.3  &   0.89 &   0.47  &  0.11 \\
\hline
\end{tabular}
\end{center}
\end{table}

\begin{figure*}
\begin{center}
\includegraphics[angle=0,width=0.80\textwidth]{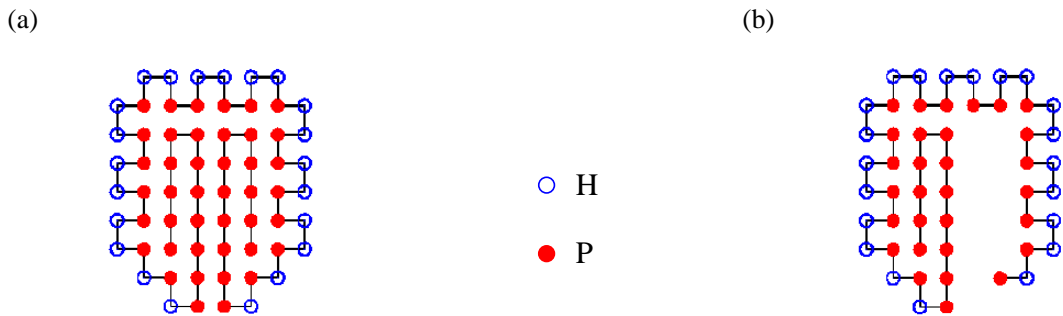}
\caption{(a) One of the ground state configurations for a $N=64$ 
chain in 2D from~\cite{Unger}.
Other states with the same energy differ in the detailed folding of the
tails in the interior, but have identical outer shapes. 
(b) When
about 3/4 of the chain is grown, one has to pass through a very
unstable configuration which is stabilized only later, when the 
hydrophobic core is filled. Adapted from Ref.~\cite{NPERM1,NPERM2}}
\label{fig-HP-64}
\end{center}
\end{figure*}

{\bf (b)} There exists one HP sequence of 64-mers introduced in 
Ref.~\cite{Unger}, for which it is particularly difficult to find
its ground state energy $E_{\rm min}=-42$ by any chain growth algorithm.
One of the ground state configuration is shown in Fig.~\ref{fig-HP-64}(a).
Its degeneracies of $E_{\rm min}=-42$ differ in the detailed folding
of the tails in the interior. As one uses a chain growth algorithm,
it seems very unnatural that the chain has to grow first along 
an arc (Fig.~\ref{fig-HP-64})(b), only until much later that the 
structure of the chain will be stabilized. It shows the difficulty
of folding this HP sequence into its ground state.
With nPERM, the ground state was reached with blind search.
The average CPU time per ground state hit was
about $30$h on the DEC21264, which seems to be roughly comparable
to the CPU time needed in Refs.~\cite{Irback,PRL831886},
but slower than Ref.~\cite{bd96} where CG was used. 
In a previous application of (old) PERM~\cite{PS321998}, the configuration
wirh $E_{\rm min}=-42$ was found only 
by means of some special tricks (non-blind search) together with the
original PERM.  

\section{DNA Melting}

At physiological temperatures, DNA forms the famous double helix. When 
the temperature is elevated, a point $T_m$ is reached where the covalent bonds
along the backbone are still strong enough to keep the two strands intact,
but the hydrogen bonds between the strands no longer can keep them together.
The ensuing separation at temperatures $>T_m$ is known as DNA denaturation
or DNA melting. 

Since the strengths of the hydrogen bonds between pairs A-T and C-G are 
different, also the melting temperatures $T_m$ are different for homogeneous
DNA, with $T_m(A-T) < T_m(C-G)$. For natural DNA, the effective melting
temperature depends on the A/C composition, and precise measurements
of melting curves for short pieces of DNA can give detailed information
about the base composition. This has been used for a long time as one 
of the easiest and fastest methods to obtain genetic information, and modern
developments have made high resolution DNA melting one of the most simple,
cheap and fast techniques for genotyping, sequence matching, and mutation
scanning \cite{Reed}.  

The sharpness of the transition in case of long homogeneous DNA has suggested 
since long ago that DNA melting is a first order phase transition 
\cite{Wartell}. But the earliest models \cite{Poland} by Poland and Scheraga
could only give rise to a second order transition, which was seen as a 
severe problem. These models of course lacked many aspects of the real 
DNA melting problem, such as the helical structure of DNA. This was done 
in view of the universality of second order phase transitions, and later 
models that did include the helix structure indeed did not do better.

On the other hand, it was already speculated early on that the excluded 
volume effect -- that was neglected in \cite{Poland} -- could be responsible
for the change into a first order transition. The first model that treated
the excluded volume effect correctly was published by Coluzzi {\it et al.}
\cite{coluzzi} and simulated by means of PERM. The model treated each DNA 
strand as a SAW on the simple cubic lattice. But the two strands were  
mutually self avoiding only to the extent that bases that were not supposed
to be bound by hydrogen bonds were not allowed to occupy the same lattice 
site. Base pairs that were bound in the native (non-molten) configuration
were not only allowed to occupy the same site, but would also gain an energy 
$\epsilon$ if they did, mimicking thereby the binding between the two strands.
In addition, variants were studied where either the excluded volume effect 
within each strand and/or between the strands was neglected.

The results were as expected: While all variants that did not incorporate
the full excluded volume effect showed second order transitions, the version
with full excluded volume interactions showed a first order transition. 
Later studies, both by simulations (using PERM and other methods) and by
analytic arguments confirmed these results (see \cite{Richard} for references).

\section{Summary}
\label{sec-Summary}
In this review we have concentrated on applications of PERM to problems
in polymer physics. But PERM can also be applied to other problems where
it is important not only to find rare events, but also to estimate
the probabilities with which they occur. This includes various
reaction-diffusion problems such as the long time tails
in the Donsker-Varadhan problem (see Sect. 3) and in the annihilation
reaction $A+A \to 0$ \cite{Mehra2}, but also to more exotic problems
like that of multiple spanning clusters in percolation
\cite{Grassberger02}. These are all problems where theory makes
clear predictions that were very hard to verify numerically with other
algorithms. But there are also some other applications of PERM to
polymer problems that we have not discussed here, such as polymers grafted to 
porous \cite{Membranes} and non-porous \cite{Graf-G05} membranes, 
adsorption of copolymers to surfaces
\cite{Ads-block}, scaling corrections for SAWs on Manhattan lattices
\cite{Manhattan}, and 2-d ISAWs with orientation dependent interactions 
\cite{Barkema}. 

PERM belongs to a class of Monte Carlo algorithms called sometimes ``sequential
algorithms with resampling" \cite{Liu}. In contrast to most other algorithms
in this class, it is implemented depth-first which leads to very compact
codes and minimal memory requirements. In principle, it can be applied to
any problem where instances are built sequentially by repeating small steps.
Its main ideas are that these steps can be biased in order to shear the
evolution towards the wanted (in general rare but highly weighted) configurations.
If this is not deemed successful, the further evolution can be pruned, while
very successful trials can be cloned, with each clone evolving further
independently. Notice that pruning and cloning are done on partially
constructed configurations, with the hope that configurations that are
successful at an early stage will also continue to be successful later.
When this is true, efficiency can be spectacular (such as for $\Theta$ polymers,
Sect. 4). But when it is not true, the method simply fails. Examples of the
latter were also discussed in Sects. 3, 4 and 11.

In most applications, the criterion for success is simply the weight of the
configuration, based on a combination of Boltzmann, entropic, and bias
compensating factors. But in some cases -- illustrated in Sect. 9 for
lattice animals -- the weight itself would be a very poor ``fitness" indicator.
For lattice animals, a much better fitness function was found empirically.

In addition to the versions of PERM that we have discussed in this review,
there exist also ``flat" \cite{flatPERM} and ``multicanonical" \cite{multiPERM}
versions of it. Their main advantage is that data over a wide range of energies
can be obtained in one single run, while ordinary PERM would need several
runs, each covering the energy range that dominates at one particular temperature.
This is certainly an attractive feature, but it is not as important as in
Markov Chain Monte Carlo algorithms. While it is there highly non-trivial
to combine results obtained at different temperatures \cite{ferrenberg},
this is much easier for PERM where the algorithm provides very precise
estimates of the partition sum.

\begin{acknowledgements}
H.-P. H. thanks K. Binder, W. Nadler, W. Paul for stimulating discussions.
She received funding from the Deutsche Forschungsgemeinschaft (DFG), grant No SFB 625/A3.
We are grateful for extensive grants of computer time at the
Cray T3E, JUMP, JUROPA and SOFTCOMP computers at the J\"ulich
Supercomputing Centre (JSC), and PC clusters at ZDV, university of Mainz.
H.-P. H. thanks K. Binder for carefully reading the manuscript.

\end{acknowledgements}

%
%



\end{document}